\documentclass[10pt, letterpaper]{article}
\newcommand{\reff}[1]{(\ref{#1})}
\newcommand{\beq}{\begin{equation}} \newcommand{\eeq}{\end{equation}}
\newcommand{\beqa}{\begin{eqnarray}}    \newcommand{\eeqa}{\end{eqnarray}}
\newcommand{\btab}{\begin{tabular}}     \newcommand{\etab}{\end{tabular}}

\newcommand{\bt}{\begin{table}}     \newcommand{\et}{\end{table}}

\newcommand{\ba}{\begin{array}}     \newcommand{\ea}{\end{array}}
\newcommand{\bc}{\begin{center}}        \newcommand{\ec}{\end{center}}
\newcommand{\bfig}{\begin{figure}}      \newcommand{\efig}{\end{figure}}
\newcommand{\bp}{\begin{picture}}       \newcommand{\ep}{\end{picture}}
\newcommand{\bq}{\begin{quote}}     \newcommand{\eq}{\end{quote}}
\newcommand{\ben}{\begin{enumerate}}    \newcommand{\een}{\end{enumerate}}

\newcommand{\mb}[1]{\ensuremath{\mathbf{#1}}}
\newcommand{\mc}[1]{\ensuremath{\mathcal{#1}}}

\newcommand{\ket}[1]{\ensuremath{\left. \left| #1  \right. \right> } }

\newcommand{\xbraket}[1]{\ensuremath{\left<  #1  \right> }}

\newcommand{\jack}{\ensuremath{J_{\lambda}(z;1/\beta)}}

\newcommand{\sm}[1]{\ensuremath{m_{(#1)}}}

\newcommand{\trace}{\ensuremath{\mbox{tr}}}
\newcommand{\Trace}{\ensuremath{ \mbox{tr}_{\Delta} } }
\newcommand{\mtext}[1]{\ensuremath{ \quad \mbox{#1}\quad } }

\begin{document}

\title{Supersymmetric Calogero-Moser-Sutherland models and Jack superpolynomials }

\author{Patrick Desrosiers\thanks{pdesrosi@phy.ulaval.ca} \\
\emph{D\'epartement de Physique},\\
Universit\'e Laval, \\
Qu\'ebec, Canada, G1K 7P4.
\and
Luc Lapointe\thanks{lapointe@scylla.math.mcgill.ca }\\
\emph{Department of Mathematics and Statistics},\\
McGill University,\\
Montr\'eal, Qu\'ebec H3A 2K6, Canada.
\and
Pierre Mathieu\thanks{pmathieu@phy.ulaval.ca - Corresponding author (address
above): tel: 418-656-3416, fax: 418-656-2040.}
\\
\emph{D\'epartement de Physique},\\
Universit\'e Laval, \\
Qu\'ebec, Canada, G1K 7P4.
}

\date{March 2001}

\maketitle

\begin{abstract}

A new generalization of the Jack polynomials that
incorporates fermionic variables
is presented. These Jack superpolynomials are constructed as those eigenfunctions of
the supersymmetric extension of the trigonometric Calogero-Moser-Sutherland (CMS)
model that decomposes triangularly in terms of  the symmetric
monomial superfunctions.  Many explicit examples are displayed.  Furthermore,
various new
results have been obtained for the supersymmetric version of the
CMS models: the Lax formulation, the construction of the Dunkl operators and
the explicit expressions for the conserved charges. The reformulation of
the models in
terms of the exchange-operator formalism is a crucial aspect of our analysis.

\end{abstract}



\newpage

\thispagestyle{plain}
 \tableofcontents

\newpage

\section{Introduction}

Calogero-Moser-Sutherland (CMS) models \cite{Calogero:1969ie,
Moser1974, Sutherland:1971ic} have been studied extensively in the
decade following their discovery. Apart from the pioneer works,
devoted mainly to the study of the energy spectrum, the
ground-state wave functions and their correlators, roughly
speaking, the initial interest of this first wave of activity was
mainly centered around their integrability and the formulation of
their various (Lie algebraic) extensions (see
\emph{e.g.},\cite{Olshanetsky:1981dk}). This subject was developed in
parallel to the soliton theory.
\footnote{Curiously, soliton equations were first defined classically  while
mechanical models were initially formulated as quantum systems.}
Although the structure of both classes of models is quite
different,
 they have some common properties, the existence of a Lax formalism
being a good example. There are however deeper and curious
connections: for instance, it has been observed that the time
evolution of the movable poles of KdV rational solutions is
governed by a dynamics of the CMS type (with a $1/r^2$ potential)
 \cite{Moser1977}.

The renewal in the interest for the CMS models that occurred in
the nineties has many sources.
One of it is rooted in the interest
for systems having fractional statistics and the  realization that the particles
subject to CMS dynamics obey fractional statistics (see for instance
\cite{Murthy1994,Bernard:1994ht,Polychronakos:1999sx} ).  This
motivation was triggered by two important problems
in condensed matter. The first  is
 the quantum Hall effect;  it has been suggested in the mid-eighties that
the quasiparticles obey fractional statistics (\emph{cf.} for instance
\cite{QHE} -- and see \cite{Azuma1994} for a relation between the
quantum Hall effect and CMS models). The early nineties brought a
second and somewhat stronger motivation in relation with high
$T_c$ superconductivity and its possible realization as a gas of
anyons (see \emph{e.g.}, \cite{Fradkin} and \cite{Ouvry1999}).
The formulation of Haldane's
generalized
 Pauli principle \cite{Haldane:1991xg} has also motivated further
theoretical considerations on the issue of fractional statistics,
this time for one-dimensional systems.

Another discovery of the late
eighties which also partly accounts for the revival of the CMS models is that
 of
 a new integrable spin-chain model with long-range interaction, the Haldane-Shastry model \cite{Haldane1988,
SriramShastry:1988gh}. This model proves to have quite remarkable
properties, among which a Yangian symmetry \cite{BernardHH}. It
turns out to be closely related to the trigonometric CMS (tCMS)
model.  Such a connection, already observed in the original
papers, has been made precise by Polychronakos in the context of
its seminal formulation of the exchange-operator formalism
\cite{Polychronakos:1992zk}.  He showed that the Haldane-Shastry
spin model can be recovered from the tCMS model augmented with
spin degrees of freedom by freezing the dynamical degrees of
freedom, thereby fixing the sites of the chains at the minima of
the tCMS potential. This observation has led to the discovery of
new integrable spin-chain models with long-range order (for
instance \cite{Fram1993, Polychronakos:1994wc}).  It has also
stimulated the interest for the study of symmetries of the
Yangian-type in these  CMS spin models. On the other hand,  the
physical motivations underlying the formulation of the
Haldane-Shastry model, related to Anderson's
resonating-valence-bond model -- at the time an alternative to the
N\'eel state in high $T_c$ superconductivity --  and the 1D
Hubbard models, have nourished the interest of CMS models in
condensed matter physics.\footnote{In that vein, we could also
mention that the Jack polynomials, the eigenfunctions of the tCMS
model -- see below -- have proved to be a rather convenient basis
for performing some computations in specific field theoretical
problems of condensed matter; see for instance \cite {Lesage}.}
Many more physical applications have been found in the last
decade, ranging from quantum chaos and matrix models
\cite{Simons1994}, mesoscopic systems \cite{Caselle1995}, black
holes \cite{Gibbons1999} to supersymmetric integrable gauge
theories \cite{D'Hoker:1999cg}.

 The return of CMS models on the hot seat in physics,  mainly through the
discoveries of many  applications, has  stimulated a wave
of activities on the models' intrinsic properties. Two objects,
discovered in quite different contexts by mathematicians, have
played a key role in that regard: these are the Dunkl operators
\cite{Dunkl1989}  and the Jack polynomials \cite{Jack1970}.

Dunkl operators play a crucial role
in various branches of mathematics, \emph{e.g.}, in
the
 theory of affine Hecke algebras
\cite{Chered} and for Schubert calculus (see \emph{e.g.},
\cite{Kirillov}). In the context of CMS models, they were a
crucial
ingredient in  the formulation of the exchange-operator
formalism, as they allowed a very simple and direct
construction of the commuting charges \cite{Minahan}. They are
also at the heart of the transfer matrix formulation of the CMS
models \cite{Bernard:1993va}.

The rebirth of CMS models coincides with a period of intense activity in
mathematics regarding the theory of symmetric functions, centered mainly on the
Jack and Macdonald polynomials,  and in particular on their combinatorial
applications.  Jack
polynomials are symmetric polynomials that were shown  to be  eigenfunctions of a
simple differential operator (see \emph{e.g.}, sect. 5 in
\cite{Stanley1988}). Then, Forrester  pointed out that this operator is precisely
the gauge transformation, by the ground-state wave function, of the tCMS
hamiltonian \cite{Forres}, bringing the subject of Jack polynomials on the
physicists' desktable.

At the time, there were still no explicit expressions for these
polynomials. The search for an explicit  description has led to
two noteworthy discoveries by physicists. The first one is a
Rodrigues' type formula, namely, a recursive
construction via the action of a differential operator built up
from shifted products of Dunkl operators \cite{Lapointe1995}.
Independently, integral formulas for the Jack polynomials have
been found in \cite{Awata:1994fz}; they led to the discovery of a
fascinating but still mysterious connection between these
polynomials and special singular vectors of the Virasoro and  W
algebras (see also \cite{Hikami:1993bc}).

On the mathematical side, explicit expressions for the Jack polynomials
have been obtained using a non-symmetric version of
these polynomials \cite{KnopSahi}.  And more recently, a rather simple
determinant formula  for the Jack polynomials has been presented
\cite{LapointeLascouxMorse}.

Some of these results -- in particular, the creation-operator formalism -- have
been extended to the construction of the hi-Jack polynomials \cite{UWadati}, which
are eigenfunctions of the CMS model with an inverse-square interaction augmented
by an harmonic confining term, and of the Macdonald polynomials
\cite{Macdonald1995,LapointeVinetM}, which are eigenfunctions of the trigonometric
Ruijsenaars-Schneider model \cite{Ruijsenaars1986},
a relativistic version of the tCMS
model.
In the latter case, integral formulas have also been obtained (see
\emph{e.g.},
\cite{Awatalivre} and refs therein).

A quite natural extension of these studies is to consider their
supersymmetric generalizations. The  supersymmetric version of the
rational CMS model has been considered by Freedman and Mende
\cite{Freedman:1990gd},
with emphasis on the study of
 supersymmetry breaking, the very physical problem that has motivated
the development of supersymmetric quantum mechanics
\cite{Witten:1981nf}.
 The
revival in the interest for the CMS models has stimulated a number of studies of
their supersymmetric counterparts
\cite{SriramShastry:1993cz, Brink:1993sz, Brink:1998zi, Wyllard:2000tm, Bordner:2000xq, Ghosh:2000pm}.  However, there remain many open problems. For
instance, there are no concise Lax formulation.
Moreover, the
suitable generalization of the Dunkl operators
is known only for the
rational model with confinement
\cite{Brink:1993sz}.
And more importantly, there are absolutely
no known results concerning the proper superextension of the Jack
polynomials. Indeed, it is only for the rational case with
harmonic term that the solutions have been constructed in
\cite{Freedman:1990gd, Ghosh:2000pm} out of
fermionic and bosonic creation operators related
to those of a supersymmetric harmonic oscillator by a similarity
transformation.

The initial goal of this work was to launch the study of the
Jack superpolynomials. As an offshoot, we have obtained a number
of new results on the supersymmetric tCMS (stCMS) models \emph{per se}.

The first step in the construction of the supersymmetric model is the
 introduction of the fermionic
variables $\theta_i$ and their conjugates $\theta_i^\dagger$.  Out of these
variables, two generic expressions for fermionic charges -- the possible
supersymmetric charges -- can be constructed. The point we want to stress
at this level is that this construction, rooted in the presence of two
fermionic charges, leads necessarily to two supersymmetries.
These charges are then used to build an hamiltonian. Explicitly, the
hamiltonian is written as the anticommutator of these two charges, which
thereby make the latter automatically conserved with respect to the dynamics
generated by this hamiltonian.\footnote{It is clear that this
supersymmetrization process is quite
different from the one used in classical field theory based on
superspace techniques. There is  no natural analogue of the
superfield here, for instance.}  We then adjust the precise expression of
the charges in order to recover, when the fermionic
variables are dropped, the bosonic hamiltonian to be supersymmetrized. The
complete hamiltonian is thus the supersymmetric hamiltonian we are looking
for. In our case, this is the stCMS hamiltonian \cite{SriramShastry:1993cz}.
This analysis is
presented in sect. 3.1

An observation that proves to be central for our subsequent
analysis is that the part of the hamiltonian that contains the
fermionic variables can be described in terms of a fermionic
exchange operator -- \emph{cf.} sect. 3.1 (and we found afterwards that
the same observation had been made before in
\cite{SriramShastry:1993cz}). This allows us to use the
projection formalism developed in \cite{Polychronakos:1992zk, Bernard:1993va} for the
description of the CMS models with spin degrees of freedom. The
key point of this projection technique is that by restricting the
space of functions on which the operators act (namely,
functions that are completely symmetric with respect to both the
fermionic and the bosonic variables), we can trade the fermionic-exchange operator -- hence, the fermionic degrees of freedom --
for a standard position-exchange operator.

In particular, this method leads us to a novel but quite natural
construction of the Dunkl operators in either their covariant or
their commuting version  -- \emph{cf.} sect. 3.3. This leads us to a
direct proof of the integrability via the construction of commuting
conserved bosonic charges. In sect. 3.2, another proof of
the integrability is presented, this one based on the Lax formalism.
Although we arrived at this Lax formulation independently, we
realized that the same Lax operators, expressed in terms of
exchange operators, had been presented in \cite{Hikami:1993bc}, albeit in a
different context.

Before pursuing the presentation of the paper's content, let us pause to
discuss briefly the meaning of integrability for supersymmetric
mechanical systems. In the non-supersymmetric case, this amounts
to demonstrate the existence of  $N$ -- the number of particles --
commuting independent bosonic charges. A working criterion for an
integrable supersymmetric extension  of an integrable mechanical
system could be the existence of $N$ commuting independent
bosonic charges that reduce to their non-supersymmetric version
when the fermionic variables are dropped. For all the cases we can
think of (including field theoretical models), this appears to be
sufficiently restrictive. However, we could argue that having
introduced $N$ new degrees of freedom (the fermionic variables being
split into a set of generalized variables, the  $\theta_i$'s, and
their conjugates,  the $\theta_i^\dagger$'s, $i=1,\dots,N$), we should expect, in
the spirit of the Liouville theorem, that $N$ additional conserved
charges are required.  Actually, the mere supersymmetry invariance
appears to supply automatically further conserved charges. As
already pointed out, the built in supersymmetry implies the
existence of  two conserved charges, denoted by $Q$ and
$Q^\dagger$. Recall that the hamiltonian is given by their
anticommutator. But this turns out to be true for all the
higher-order
 hamiltonians of the system, \emph{i.e.}, they can all be expressed  as
anticommutators of higher-order fermionic charges. Indeed, for the
stCMS model, we can construct rather directly (using the Dunkl
operators, for instance) $2N$ conserved fermionic charges.
However, it should be stressed that these do not anticommute among
themselves. Moreover, by inspection, we readily find $N$
additional bosonic conserved charges that commute with the bosonic
ones previously constructed. Afterwards, this appears to be
somewhat natural given that we have two supersymmetries,
suggesting heuristically that the charges get organized in
`multiplets' of four, two bosonic and two
fermionic.\footnote{This can be compared to the case of the
classical ${\cal N}=1$ supersymmetric Korteweg-de Vries equation
\cite{Dargis:1993jf}, which is  probably the best studied supersymmetric
integrable system. In that case, we find that in addition to the bosonic
supersymmetric extension of the usual KdV charges, there are
(twice as many) nonlocal fermionic charges whose Poisson brackets
yield a local bosonic charge - if dimensionally allowed - and
vanish otherwise.}

In section 4 we turn to the main subject of this work: the
formulation of the Jack superpolynomials. They are defined as
eigenfunctions of the stCMS model. Notice that by a superpolynomial we refer to
a polynomial in bosonic and fermionic variables without imposing a
supersymmetric invariance constraint (\emph{i.e.}, these are not
supersymmetric polynomials). We first unravel, in sect. 4.1, the
mixed symmetry properties, with respect to the bosonic variables,
that are induced by the presence of the fermionic variables on any
symmetric superpolynomials. This leads us naturally to the central
concept of superpartitions introduced in sect. 4.2 and which
appears to be original. Superpartitions are used in turn to define
the monomial superfunctions. Jack superpolynomials are then
defined in sect. 4.3 as those stCMS eigenfunctions that are
triangular with respect to a monomial superfunction decomposition.
Many examples are presented.

Various straightforward extensions of the results presented in
this paper, directions for future research and conclusions are
collected in the final section. Some auxiliary sections
complete the article.  A brief review of the basic definitions
pertaining to the usual Jack polynomials and some associated
concepts is presented in sect. 2. The remaining complementary
material is spread in
 three appendices. Excited states can be built from a vacuum state free of
fermions, as it is done in the main body of the paper, or from a `vacuum' filled
by $N$ fermions. This second option is considered in app. A. In app. B, we
introduced creation operators analogous to those introduced in \cite{Lapointe1995} for
the standard Jack polynomials. Finally, a simple combinatorial expression
counting the number of superpartitions of a given degree and a given fermionic
number is presented in app. C.

\section{Background}

\subsection{Calogero-Moser-Sutherland models}

The CMS-models describe systems of $N$
particles interacting pairwise through long-range potentials. The
classical and quantum versions of these models are integrable.  In
this article, we focus on the supersymmetric extension of the
 quantum tCMS model in
which the identical particles of mass $m$ lie on a circle of
circumference $L$.  If we set $m=\hbar=1$, the hamiltonian of the
tCMS model is the following \cite{Sutherland:1971ic}:
\beq \label{hamii}
H=\frac{1}{2}\sum_{i=1}^{N}p_i^2+\left(\frac{\pi}{L}\right)^2\beta(\beta-1)\sum_{1\leq
i<j\leq N}\frac{1}{\sin^2(\pi x_{ij}/L)}, \eeq where $\beta$ is a
dimensionless real coupling constant. In this equation, and for
the remainder of the article, double indicing stands for the
difference between two variables, \emph{i.e.}, \beq x_{ij} \equiv
x_i-x_j. \eeq Position and momentum variables obey the usual
commutation relations: \beq [x_j,p_k]=i\delta_{jk}. \eeq
Two other models can be obtained from the
tCMS model: the replacement
 $L \rightarrow iL$ yields the hyperbolic model,
whereas the limit $L\rightarrow \infty$ gives the rational model (on
an infinite line).

The hamiltonian \reff{hamii} is semi-positive \emph{i.e.}, \beq
H=\frac{1}{2}\sum_j A_j^\dagger A_j+E_0 \, , \eeq with \beq
A_j=p_j-i\sum_{k\neq j} X_{jk} \, .\eeq
 Hence, the minimal value in the spectrum of $H$ is given by
\beq \label{evide} E_0=\left(\frac{\pi
\beta}{L}\right)^2\frac{N(N^2-1)}{6}. \eeq The ground state, which
is annihilated by every operator $A_j$, corresponds to the
following Jastrow-type function: \beq \label{videsCMS}
\psi_0(x)=e^{\sum_{j<k}\int dx_j
X_{jk}}=\prod_{j<k}\sin^\beta\left(\frac{\pi
x_{jk}}{L}\right)\equiv \Delta^\beta (x)\, . \eeq

The simplest way of showing the integrability of the quantum CMS
models is by displaying  a Lax pair, namely
  two $N
\times N$ Hermitian matrices, denoted by $L$ and $M$, satisfying the
relations:
\beq
\dot{L}_{jk}=-i[L_{jk}, H]=-i[L,M]_{jk} \quad
\mbox{and} \quad \sum_j M_{jk}=\sum_k M_{jk}=0 \, .
\label{defpaireLax}
 \eeq
The constraint on $M$ is essential for
the following $N$ independent quantities to be conserved:
\beq\label{quantcons}
H_{(n)}=\frac{1}{n} \Trace L^n \equiv
\frac{1}{n} \trace (L^n \mb{\Delta}), \eeq where $\mb{\Delta}$ is
the matrix whose entries are all $1$'s. Therefore, $\Trace A $
denotes the `total trace' of $A$, that is,  the sum of all the
entries of $A$.  For the tCMS model, the Lax pair
reads \cite{Moser1974}:
\beq \label{defLax} \ba{c}
L_{jk}=p_j\delta_{jk}+i(1-\delta_{jk})X_{jk}\, ,\\
M_{jk}=\delta_{jk}\sum_{l\neq j}X'_{jl}-(1-\delta_{jk})X'_{jk},
\ea
\eeq
where $X'_{jk}=dX(x_{jk})/dx_{jk}$.  From eq. \reff{quantcons}, we see that
the first and second conserved quantities correspond respectively to the
momentum $P=\sum_i p_i$ and the
hamiltonian $H$ of the system.

In order to solve the Schr\"odinger equation associated to the CMS model, it is
convenient to set:
\beq
z_j=e^{2\pi i x_j/L} \, .
\eeq
The variable $z_j$ thus gives the  position of the $j^{th}$ particle on a circle
of circumference $L$ in the complex plane.  In this notation,
$H$ becomes:
\beq \label{hnonjack}
H=2 \left(\frac{\pi}{L}\right)^2\left[\sum_i \left(z_i \frac{\partial}{\partial z_i}\right)^2-2\beta(\beta-1)\sum_{i<j}\frac{z_i z_j}{z_{ij}^2}\right].
\eeq
The eigenfunctions of the excited states of the hamiltonian \reff{hnonjack} are
written in the form $\psi(x)=\phi(x)
\psi_0(x)$ where $\phi(x)$ is required to be symmetric in order for $\psi$ to
behave like $\psi_0$ under the exchange of particles.
It is thus natural to conjugate the
hamiltonian with the ground-state
wave function $\psi_0(x)$:
\beq \label{hjack}
\ba{rcl}
\bar{H}&=& \frac{1}{2} \left(\frac{L}{\pi}\right)^2 \Delta^{-\beta}(H
-E_0)\Delta^{\beta} \, ,\\
 &=& \sum_i (z_i \partial_i)^2+\beta \sum_{i<j}\frac{z_i+z_j}{z_{ij}}(z_i \partial_i-z_j\partial_j)\, ,
\ea
\eeq
and look for the eigenfunctions $\phi(x)$ of this conjugated hamiltonian.
In the following, `bar' operators will stand for operators that have
been obtained by a similar conjugation of the ground state.

The symmetric eigenfunctions $\phi(x)$ of \reff{hjack} are known as
the Jack polynomials.

\subsection{Symmetric functions and Jack polynomials}

We now  summarize some basic results concerning
 symmetric
functions \cite{Macdonald1995}.  This will allow us to define
properly the Jack polynomials \cite{Jack1970, Stanley1988},
and thereby, to present the solutions of the tCMS model
\cite{Sutherland:1971ic}.

\noindent \emph{1- Symmetric functions and exchange operators}. Symmetric
functions are invariant under the action of the symmetric (or permutation)
group $S_N$. If $z=(z_1, \ldots, z_N)$ denotes the set of variables, then a
function $\mc{F}(z)$ is symmetric if it remains invariant under the exchange of
its variables:
\beq
    K_{ij}\mc{F}(z)=\mc{F}(z) \quad \forall i,j\, ,
\eeq
where  $K_{ij}$ is a transposition of $S_N$, \emph{i.e.}, an exchange operator,
whose action is defined as follows:
\beq
        K_{ij}f(z_i, z_j)=f(z_j, z_i,)K_{ij} \, ,
\eeq
with  $f(z_i,z_j)$ standing for a function or
an operator. The fundamental properties of the exchange operators are
\beq \label{proprioK}
    K_{ij}=K_{ji} \mtext{,}
    K_{ij}^\dagger=K_{ij}\mtext{,}
    K_{ij}K_{jk}=K_{ik}K_{ij}=K_{jk}K_{ki}\mtext{,}
    K_{ij}^2=1.
\eeq
Let  $\Sigma_N$ denote the ring of symmetric polynomials in the variables
$z_1,\dots,z_N$, and let $\Sigma_N^{(n)}$ be
the subspace of
symmetric polynomials of degree $n$.  Before introducing various bases
for
$\Sigma_N$, we need to introduce
the notion of
partitions.

\noindent \emph{2- Partitions}.  A partition is a weakly-decreasing sequence of
non-negative integers.  More precisely, a  partition $\lambda$ of weight (or
degree) $n$ is defined as follows:
\beqa
&\lambda = (\lambda_1,\lambda_2, \dots, \lambda_l)\, ,&\nonumber\\
&\lambda_1 \geq \lambda_2 \geq \dots \geq \lambda_l \geq 1\, ,&
\nonumber\\
&n=\lambda_1+
\lambda_2 + \dots \lambda_l=|\lambda|\, ,&
\eeqa
where $l=l(\lambda)$ is the length of the partition, that is, the number of
its non-zero parts.     We use $p(n)$ for the
number of partitions of $n$, \emph{e.g.}, the number of partitions of 4 is
$p(4)=5$.  We can also represent a partition $\lambda$ as:
\beq
\lambda=(1^{m_1}, 2^{m_2},\ldots, i^{m_i}, \ldots)\, ,
\eeq
where $m_i$ is the number of parts of $\lambda$ equal to $i$.
This allows
us  to define the following constant
\begin{equation}\label{zlambda}
z_{\lambda}
    = 1^{m_1} m_1!~ 2^{m_2} m_2! \dots \, ,
\end{equation}
which
enters in the definition of the Jack polynomials.
There exists a natural partial order on partitions called the dominance ordering.
It is defined in the following
way:
\beq
     \lambda \geq \mu \mtext{if}
    \lambda_1+\lambda_2+\dots
+
\lambda_i \geq \mu_1 + \mu_2 +\dots +\mu_i \, , \quad \forall i \, .
\eeq
The dominance ordering is a total order only for weights up to $n=5$.
Note finally that to $\lambda$ we can associate a Young tableau having
$\lambda_i$ boxes in the $i$-th row. The conjugate partition, denoted $\lambda'$
corresponds to the partition resulting from the interchange of the rows and
columns in the Young tableau associated to $\lambda$.

\noindent \emph{3- Power sums}.  The symmetric polynomials
\beq
    p_n=\sum_i z_i^n,
\eeq
where the sum extends over the $N$ variables, are called power sums.
The set of all products of power sums, \emph{i.e.},
\beq
p_{\lambda}
    =p_{\lambda_1}p_{\lambda_2}\dots p_{\lambda_l},
\eeq
forms a basis of $\Sigma_N$.

\noindent \emph{4- Elementary symmetric functions}.  The elementary symmetric
functions are:
\beq
    e_n = \sum_{i_1<i_2<\ldots<i_n} z_{i_1}\ldots z_{i_n}.
\eeq
Again, the set of all products of elementary functions
\beq
e_\lambda=e_{\lambda_1}\ldots e_{\lambda_l} \, ,
\eeq
is a basis of $\Sigma_N$.

\noindent \emph{5- Monomial symmetric functions}.  The monomial symmetric
functions are
 defined as follows:
\beq
m_{\lambda}
    = {\sum_{P\in S_N}}' z^{P(\lambda)}={\sum_{P\in S_N}}' z_1^{\lambda_{P(1)}}
z_2^{\lambda_{P(2)}}\cdots z_N^{\lambda_{P(N)}} \, ,
\eeq
where  here and below, the prime on the sum is used to indicate that it is done
only over distinct permutations, which means that no monomial is  repeated.
The
$p(n)$ possible monomial functions of degree
$n$ constitute another basis of $\Sigma_N^{(n)}$.  Also, the monomial symmetric
functions generalize $e_n$ and $p_n$:
\beq
m_{(n)}=p_n \mtext{and} m_{(1^n)} = e_n .
\eeq
The simplest
monomial functions are given in Table \ref{monomiales4}.

\bt
\caption{Monomial functions of weight $|\lambda| \leq 3$ for $N=4$ variables}
\bc
\label{monomiales4}
\btab{|c||c|c|}
\hline
Weight & Partition &Monomial function \\
$|\lambda|$&$\lambda$&$m_{\lambda}(z)$\\
\hline\hline
0& (0) & 1\\ \hline
1& (1) & $z_1+z_2+z_3+z_4 $\\ \hline
2& (11)& $z_1 z_2+z_1 z_3+z_1 z_4+z_2 z_3 $\\ & &$+z_2 z_4+z_3 z_4$\\
 &(2)&$z_1^2+z_2^2+z_3^2+z_4^2$\\ \hline
3&(111)&$z_1 z_2 z_3+z_1 z_2 z_4+z_1 z_3 z_4+z_2 z_3 z_4$\\
 &(21)&$z_1^2 z_2+z_1^2 z_3+z_1^2 z_4+z_2^2 z_1$\\& &$+z_2^2 z_3+z_2^2 z_4+z_3^2 z_1+z_3^2 z_2+z_3^2 z_4$\\& &$+z_4^2 z_1+z_4^2 z_2+z_4^2 z_3$\\
 &(3)&$z_1^3+z_2^3+z_3^3+z_4^3$\\ \hline
\etab
\ec
\et

\noindent \emph{6- Jack polynomials}. The Jack polynomials,
$J_{\lambda}(z_1,\dots,z_N;\alpha)$, are symmetric polynomials depending on a
parameter $\alpha$ that also
 form a basis of $\Sigma_N$. They belong to the ring
$\mb{Q}(\alpha)[z_1, \ldots, z_N]_{S_N}$ of symmetric polynomials with rational
coefficients in  $\alpha$.  They are
uniquely characterized by the following two
conditions (see \emph{e.g.},\cite{Macdonald1995}):
\beqa
\label{cond1}
\xbraket{J_{\lambda},J_{\mu}}_\alpha
    &=& 0  \mtext{and} \lambda \neq \mu \quad \mbox{ (orthogonality)}\, ,\\
\label{cond2}
J_{\lambda}\left(z;  \alpha \right)
    &=& m_{\lambda}+\sum_{\mu < \lambda} v_{ \lambda \mu}(\alpha)
m_{\mu}\quad \mbox{ (unitriangularity)}\, ,
\eeqa
where the scalar product is defined in the following way,
with respect to the power sums:
\begin{equation}
\langle p_{\lambda}, p_{\mu} \rangle_\alpha
    = \delta_{\lambda,\mu} z_{\lambda}\alpha^{l(\lambda)}\, ,
\end{equation}
where $z_{\lambda}$ has been introduced in eq. \reff{zlambda}.
The Jack polynomials generalize several types of symmetric polynomials:
\beq
 J_\lambda(z;\alpha)\rightarrow
 \left\{ \ba{l@{\quad} l@{\quad , \quad }l}
s_\lambda(z)&\mbox{(Schur functions)}& \alpha\rightarrow 1\\
m_\lambda(z)&\mbox{(monomial functions)}& \alpha \rightarrow \infty \\
e_{\lambda^\prime}(z)&\mbox{(elementary functions)}& \alpha
\rightarrow 0 \ea \right. \eeq
(recall that $ \lambda^\prime$ refers to the conjugate partition).

A few examples of Jack polynomials expanded in terms of monomial
symmetric functions are shown in Table \ref{tabjacks}.  In this
notation, the number of variables is irrelevant as long as it is
not smaller than the degree of the polynomial.

\bt
\caption{Jack
polynomials of weight $|\lambda|\leq 4$}
 \bc \label{tabjacks}
\btab{|c||c|c|c|}
\hline
Weight & Partition &Eigenvalue& Jack polynomials\\
$|\lambda|$&$\lambda$&$\varepsilon_{\lambda}(\beta,N)$&$J_{\lambda}(z;1/\beta)$\\
\hline\hline
0& $(0)$ &0&$ m_{(0)}$\\ \hline
1& $(1) $&$1+\beta N-\beta$& $m_{(1)} $\\ \hline
2& $(1^2)$ &$2+2\beta N-4\beta$ & $m_{(1^2)} $\\
 & $(2) $&$4+2\beta N-2\beta$ & $m_{(2)} +\frac{2 \beta}{1+\beta}m_{(1^2)}$\\ \hline
3&$(1^3)$ & $ 3+3\beta N-9\beta $ & $ m_{(1^3)}$ \\
 &$(21)$ & $ 5+3\beta N-5\beta $ & $m_{(21)}+\frac{6 \beta}{1+2\beta}m_{(1^3)} $ \\
 &$(3)$ & $ 9+3\beta N-3\beta $ & $m_{(3)}+\frac{3\beta^2}{2+\beta}m_{(21)}+\frac{6\beta^2}{(1+\beta)(2+\beta)}m_{(1^3)} $\\
\hline
4&$(1^4)$ & $ 4+4\beta N-16\beta $ & $m_{(1^4)} $ \\
 &$(21^2)$ & $ 6+4\beta N-10\beta$ & $m_{(21^2)}+\frac{12\beta}{1+3\beta}\sm{1^4} $ \\
 &$(2^2)$ & $ 8+4\beta N-8\beta $ & $\sm{2^2}+\frac{2\beta}{1+\beta}\sm{21^2}+\frac{12\beta^2}{(1+\beta)(1+2\beta)}\sm{1^4} $ \\
  & $(31)$ & $ 10+4\beta N-6\beta $ & $\sm{31}+\frac{2\beta}{1+\beta}\sm{2^2}+\frac{\beta(3+5\beta)}{(1+\beta)^2}\sm{21^2}+\frac{12\beta^2}{(1+\beta)^2}\sm{1^4} $ \\
 &$(4)$ & $ 16+4\beta N-4\beta $ & $ \sm{4}+\frac{4\beta}{3+\beta}\sm{31}+\frac{6\beta(1+\beta)}{(2+\beta)(3+\beta)}\sm{2^2}$\\
 & & & $+\frac{12\beta^2}{(2+\beta)(3+\beta)}\sm{21^2}+\frac{24\beta^3}{(1+\beta)(2+\beta)(3+\beta)}\sm{1^4} $ \\
\hline
\etab
\ec
\et

\noindent \emph{7- Jack polynomials and the tCSM model}. Jack
 polynomials are related to the tCMS model for:
\beq
\alpha\equiv 1/ \beta
\eeq
where $\beta$ is the model's coupling constant.
In this context, we
have to replace the scalar product  $\xbraket{A,B}_\alpha$ by the physical
one:
\beqa \label{physscal}
    \xbraket{A(x),B(x)}&=&\oint \frac{dz_1}{2\pi i}\ldots \frac{dz_N}{2\pi i}\prod_{i\neq j}\left(1-\frac{z_i}{z_j}\right)^\beta A(1/z) B(z)\, ,\nonumber\\
& \propto &\int_0^{2\pi}dx_1\ldots dx_N|\psi_0|^2 A(x)^{\ast} B(x)\, ,
\eeqa
where $\psi_0$ is the ground state of the trigonometric model.  The hamiltonian
$\bar{H}$ defined in \reff{hjack} is self-adjoint with respect to the scalar
product \reff{physscal}.
This physical scalar product can equally be used to characterize
the Jack polynomials as the unique polynomials satisfying \reff{cond2} that
are orthogonal
with respect to \reff{physscal}.

It is also known that the Jack polynomials are eigenfunctions
of the transformed hamiltonian $\bar{H}$ \cite{Stanley1988, Forres}:
\beq
 \bar{H} J_\lambda (z;1/\beta)= \varepsilon_\lambda J_\lambda (z;1/\beta)
\eeq
with eigenvalues:
\beq
\varepsilon_\lambda=\sum_j [\lambda_j^2+\beta(N+1-2j)\lambda_j].
\eeq
The wave functions of the original trigonometric model are now simply
 $\psi_\lambda(z)=J_\lambda (z;1/\beta) \Delta^\beta$, with eigenvalues
$E_\lambda=2(\pi/L)^2\varepsilon_\lambda+E_0$.  Therefore, if we
introduce the quasi-momenta \beq \kappa_i=\left(\frac{2
\pi}{L}\right)[ \lambda_i +\beta(N+1-2i)], \eeq we observe that
the spectrum of the model is that of a system of $N$ free
quasi-particles, each of these with quasi-momentum $\kappa_i$ :
\beq E_\lambda=\sum_i \frac{\kappa_i^2}{2}. \eeq The quasi-momenta
of two neighboring quasi-particles satisfies: \beq
\kappa_i-\kappa_{i+1}\geq \frac{4\pi \beta}{L}. \eeq The excited
states of the tCMS model thus obey a generalized
exclusion principle \cite{Haldane:1991xg,Bernard:1994ht, Polychronakos:1999sx}. In particular, we recover free bosons
if $\beta=0$ and free fermions if $\beta=1$.

\section{Supersymmetric Calogero-Moser-Sutherland models}

\subsection{Supersymmetric quantum mechanics}

Consider a quantum model that contains both bosonic and fermionic
variables and whose hamiltonian is denoted  $\mc{H}$.
Following the usual
methods of supersymmetric quantum mechanics \cite{Nicolai1976, Witten:1981nf, Cooper:1995eh}, we consider, in
addition to the $2N$ bosonic variables $(x,p)$, the $2N$ fermionic
variables  $(\theta,\theta^\dagger)$.\footnote{This amounts to considering
 $\mc{N}=2$
supersymmetries -- \emph{i.e.}, there will be two conserved supersymmetric
charges. Extensions to more supersymmetries are discussed in the conclusion.}
 The bosonic
and fermionic variables
satisfy respectively a Heisenberg and a Clifford
algebra:
\beq
 [x_j,p_k]=i \delta_{jk} \quad,\quad\{\theta_j, \theta_k^\dagger\}=\delta_{jk}, \label{algquant}
\eeq
with all other commutators or anticommutators equal to zero.  We will usually
work with a differential realization of these algebras:
\beq
p_j=-i \frac{\partial}{\partial x_j} ,\quad \theta_j^\dagger=\frac{\partial}{\partial \theta_i}. \label{diffrepre}
\eeq

To construct a supersymmetric hamiltonian, we will first construct  two
supersymmetric charges, denoted
$Q$ and $Q^\dagger$, and define the
 hamiltonian  as their anticommutator:
\beq
    \mc{H}=\frac{1}{2}\{Q, Q^\dagger\} \, .
\eeq
By construction, the hamiltonian's eigenvalues are non-negative.
The hamiltonian is invariant under a supersymmetric transformation if:
\beq
    Q^2=(Q^\dagger)^2=0. \label{consusyq}
\eeq
 By writing the charges under the form
 \beq Q=\sum_{i=1}^N
\theta^\dagger_i A_i(x,p) ,\quad Q^\dagger =\sum_{i=1}^N
\theta_i A^\dagger_i(x,p)\, ,
 \eeq
we find that eq. \reff{consusyq} requires:
\beq [A_i,A_j]=0=[A_i^\dagger,A_j^\dagger] ,\quad\forall i,j. \eeq
The generic supersymmetric hamiltonian is thus:
\beq
\mc{H}=\frac{1}{2}\left(\sum_i A_i^\dagger A_i
+\sum_{i,j}\theta_i^\dagger\theta_j [A_i,A_j^\dagger] \right)\, .
\label{defhgen}
\eeq
 For non-relativistic models, the hamiltonian
is proportional to the square of the particles' speed.  We
therefore write $A_i$
as a linear function of the momentum $p_i$: \beq \label{defQ}
Q=\sum_j \theta_j^\dagger(p_j-i\Phi_j(x)), \quad Q^\dagger=\sum_j
\theta_j(p_j+i\Phi_j(x))  \label{constcharge} \eeq From eq. \reff{consusyq},
the potential $\Phi_j(x)$ must be of the form
\beq \label{prepot}
\Phi_j(x)=\partial_{x_j}W(x)\,,
\eeq
 where $W(x)$
(called the prepotential), is an arbitrary
function of the variables $x_1,\dots,x_N$. The supersymmetric
hamiltonian now takes the form \cite{Freedman:1990gd}:
\beq
\mc{H}=\frac{1}{2}\sum_{i}(p_i^2+(\partial_{x_i}W)^2+\partial^2_{x_i}W)-\sum_{i,j}\theta_i\theta_j^\dagger\partial_{x_i}\partial_{x_j}W\, .
\eeq This hamiltonian is an extension of the purely bosonic model
whose potential is
 $\sum_i[(\partial_{x_i}W)^2+\partial^2_{x_i}W]$.

Since the hamiltonian is semi-positive, any state annihilated by the charges
$Q$ and $Q^\dagger$ is
a ground state (vacuum). Obviously, only the
vacuum is supersymmetric since
an excited state cannot be simultaneously annihilated by
both charges. The charges defined in
\reff{constcharge} naturally lead to two ground states:
\beq
\psi_0=e^{W}\ket{0}\mtext{and} \tilde{\psi}_0=e^{-W}\ket{\tilde{0}}, \label{viedesusy}
\eeq
where the ground states $\ket{0}$ and $\ket{\tilde{0}} $ belong to the fermionic
Fock space and are defined as follows:
\beq
\theta_i \ket{\tilde{0}} =0, \quad \theta_i^\dagger \ket{0}=0, \quad \forall i.
\eeq
If we interpret the $\theta_i$'s as fermionic creation operators, then
$\ket{\tilde{0}} $ and $\ket{0}$ correspond to the $N$-fermion and the 0-fermion
states respectively.  In the realization
\reff{diffrepre}, the ground states must then be of the form:
\beq
 \ket{0}\rightarrow 1, \quad \ket{\tilde{0}}\rightarrow \theta_1 \ldots \theta_N.
\eeq
To be physically meaningful, the functions $\psi_0$ and/or
$\tilde{\psi}_0$ must be normalizable.  If this is not the case,
the supersymmetry is said to be broken.  It should be noted
that eq. \reff{viedesusy} provides a natural way to supersymmetrize a model.
Knowing the ground state
$\psi_0$ of that model, it suffices to let $W=\ln \psi_0$ to get its
supersymmetric extension \cite{Cooper:1995eh}.

We now specialize to a  prepotential
of the form: \beq W(x)=\sum_{i<j}w(x_{ij})\, .
\eeq
The comparison of eqs \reff{videsCMS} and \reff{viedesusy}
immediately gives the right choice of $W$
 for the CMS models: \beq
\label{defW} w'(x_{ij})=X_{ij}. \eeq Consequently, the
stCMS hamiltonian
reads: \beqa
\mc{H}&=&\frac{1}{2}\sum_{i} p_i^2
+\sum_{i < j}[X_{ij}^2+X'_{ij}(1-\theta_{ij}
\theta^\dagger_{ij})]-N(N-1)(N-2)\left(\frac{\pi\beta}{L}\right)^2 \, ,\nonumber\\
&=&\frac{1}{2}\sum_{i=1}^{N}p_i^2+\left(\frac{\pi}{L}\right)^2\sum_{
i<j}\frac{\beta(\beta-1+\theta_{ij} \theta^\dagger_{ij})}{\sin^2(\pi
x_{ij}/L)}-E_0,
\eeqa
where $E_0$ is as given in \reff{evide}.  The two ground states
\beq
\psi_0(x) =\Delta^\beta(x) \mtext{and} \tilde{\psi}_0(x,\theta)
=\Delta^{-\beta} (x)\theta_1\cdots\theta_N
\eeq
are invariant under
supersymmetric transformations and  are normalizable for any value of
$\beta$.

The supersymmetric model can be solved much more easily if
we notice that the term
\beq
\kappa_{ij}\equiv 1-\theta_{ij}\theta^\dagger_{ij}= 1-(\theta_{i}-\theta_j)(\partial_{\theta_i}-\partial_{\theta_j}).
\eeq
is a fermionic-exchange operator \cite{SriramShastry:1993cz}, that is,
\beq
\kappa_{ij}\,
f(\theta_i,\theta_j,\theta^\dagger_i,\theta^\dagger_j)=
f(\theta_j,\theta_i,\theta^\dagger_j,\theta^\dagger_i)\, \kappa_{ij}
\eeq
for any monomial function $f$.  Moreover, the
$\kappa_{ij}$'s satisfy the usual properties
 \reff{proprioK} of exchange
operators.

As in the non-supersymmetric case, it is
convenient to use
$z_j=e^{2\pi i x_j/L}$,
in terms of which the hamiltonian reads:
\beq
\mc{H}=2 \left(\frac{\pi}{L}\right)^2\left
[\sum_i \left(z_i \partial_i\right)^2-2\sum_{i<j}\frac{z_i
z_j}{z_{ij}^2}\beta(\beta-\kappa_{ij})\right]-E_0\, .
\eeq
Removing the ground-state contribution leads to:
\beq \label{shjack}
\ba{rcl}
\bar{\mc{H}}&\equiv& \frac{1}{2} \left(\frac{L}{\pi}\right)^2 \Delta^{-\beta}\mc{H}\Delta^{\beta}\\
 &=& \sum_i (z_i \partial_i)^2+\beta \sum_{i<j}\frac{z_i+z_j}{z_{ij}}(z_i \partial_i-z_j\partial_j)-2\beta\sum_{i<j}\frac{z_i z_j}{z_{ij}^2}(1-\kappa_{ij}) \, ,
\ea
\eeq
which is  still supersymmetric because it is invariant under the
action of the transformed fermionic charges:
\beq
\bar{Q}=\Delta^{-\beta} Q\Delta^\beta \mtext{and}
\bar{Q}^\dagger=\Delta^{-\beta} Q^\dagger\Delta^\beta.
\eeq
A complete set of eigenfunctions of the hamiltonian
 \reff{shjack} is given in sect. \ref{superjack}.
Excited states built from the  second ground state are considered in appendix
\ref{etatsantisym}.

\subsection{Lax formalism in the supersymmetric CMS models}

Quite remarkably, knowing the CMS models' Lax pair is enough to
guarantee the existence of their supersymmetric extensions.
Moreover, the supersymmetric Lax pair is a simple
extension of the non-supersymmetric one.

The first statement is proved as follows. Recall that the supersymmetric
hamiltonian is the anticommutator of the two supersymmetric charges.
Comparing eqs \reff{defLax}, \reff{defQ}, \reff{prepot} and \reff{defW}, we see
that the supersymmetric charges, hence the supersymmetric hamiltonian, can
easily be built from the Lax matrices \cite{SriramShastry:1993cz}:
\beq
\ba{l}
Q=\sum_{i,j} \theta^\dagger_j L_{ij} \, ,\\
Q^\dagger=\sum_{i,j} \theta_i L_{ij}\, .\\
\ea
\eeq
Therefore, the Lax matrices of the quantum CMS models guarantee the
existence of their supersymmetric extensions!

In order to prove the integrability of the supersymmetric models, we introduce
the four  matrices
that will provide the Lax formulation of the supersymmetric system:
\beq
\ba{l}
\mc{L}_{jk}=p_j\delta_{jk}+i(1-\delta_{jk})X_{jk}\kappa_{jk}\, ,\\
\mc{M}_{jk}=\delta_{jk}\sum_{l\neq j}X'_{jl}\kappa_{jl}-(1-\delta_{jk})X'_{jk}\kappa_{jk}\, .\\
\Theta_{jk}=\theta_j\delta_{jk}\, ,\\
\Theta^\dagger_{jk}=\theta^\dagger_j\delta_{jk}\, .\\
\ea \eeq
These matrices  obey the relations: \beq \ba{l}
\dot{\mc{L}}_{jk}=-i[\mc{L}_{jk}, \mc{H}]=-i[\mc{L},\mc{M}]_{jk}\, ,\\
\dot{\Theta}_{jk}=-i[\Theta_{jk}, \mc{H}]=-i[\Theta,\mc{M}]_{jk}\, ,\\
\dot{\Theta}^\dagger_{jk}=-i[\Theta^\dagger_{jk}, \mc{H}]=-i[\Theta^\dagger,\mc{M}]_{jk}\, .\\
\ea
\eeq
To verify the  equivalence between theses relations  and the
equations of motion, we use the properties
\beq
\mc{L}_{jk}\theta_k=\theta_j\mc{L}_{jk}\mtext{,}\mc{L}_{jk}\theta^\dagger_k=\theta^\dagger_j\mc{L}_{jk}
\eeq
 and
\beq \sum_i\mc{M}_{ij}=\sum_j\mc{M}_{ij}=0\Leftrightarrow
\mb{\Delta}\mc{M}=\mc{M}\mb{\Delta}=0, \eeq where
$\mb{\Delta}_{ij}=1$.

We see that $\mc{L}$ and $\mc{M}$  are obtained from $L$ and $M$
by simply multiplying  each $X_{jk}$ or $X'_{jk}$ factor by the exchange
operator $\kappa_{jk}$.

Using the matrices  $\mc{L}$ and ${\Theta}$, one can construct
the following independent quantities that can easily be shown to be conserved:
\beq
\ba{l l}
\mc{H}_{(n)}=\frac{1}{n}\Trace \mc{L}^n=\frac{1}{n}\sum_{jk} \mc{L}^n_{jk}, & n=1, 2, \ldots, N \, ,\\
\mc{Q}_{(n)}=\frac{1}{n}\Trace (\Theta \mc{L}^n)=\frac{1}{n}\sum_{jk} \theta_{j} \mc{L}^n_{jk}, & n=0,1, \ldots, N-1\, ,\\
\mc{Q}^\dagger_{(n)}=\frac{1}{n}\Trace (\Theta^\dagger \mc{L}^n)=\frac{1}{n}\sum_{jk} \theta^\dagger_{j} \mc{L}^n_{jk}, & n=0,1, \ldots, N-1\, ,\\
\mc{I}_{(n)}=\frac{1}{n}\Trace (\Theta \Theta^\dagger \mc{L}^n)=\frac{1}{n}\sum_{jk} \theta_j\theta^\dagger_{j} \mc{L}^n_{jk}, & n=0,1, \ldots, N-1\, .\\
\ea
\eeq
More generally, any operator that can be written as the total trace of a
polynomial function $F$ only depending on the matrices  $\Theta, \Theta^\dagger$
and $\mc{L}$ is conserved:
\beq
\frac{d}{dt} \Trace F(\Theta, \Theta^\dagger, \mc{L})=0\, .
\eeq
The quantities $\mc{Q}_{(1)}$ and $\mc{Q}_{(1)}^\dagger$
are simply the generators of the supersymmetric transformations.
However, the fermionic charges are not in involution:
their
 anticommutators generate the hamiltonians
 $\mc{H}_{(n)}$ (see below).

We stress that the results of this subsection apply
to all types of supersymmetric
CMS models and not just the stCMS one.

\subsection{Dunkl operator formalism and the supersymmetric
Calogero-Moser-Sutherland models}\label{SUSYDunkl}

In this section we construct the Dunkl operators of the stCMS
model. With these operators in hands, the integrability of the
model can be very easily (re)established.

We first introduce  a new exchange operator that
acts on the bosonic and fermionic variables:
\beq
\mc{K}_{ij}\equiv \kappa_{ij}K_{ij}\, , \mtext{where} [\kappa_{ij},K_{ij}]=0.
\eeq
A function of the variables
$z_i$ and $\theta_i$ is said to be a symmetric superfunction if it is invariant
under the action of the
$\mc{K}_{ij}$'s.  It is worth noticing that
the action of
$\kappa_{ij}$  on symmetric superfunctions is equivalent to the action of
 $K_{ij}$ on those functions.  For instance, if $\mc{F}_{\mc{K}}$
is a symmetric superfunction, that is,
\beq
\kappa_{ij}\mc{F}_{\mc{K}}=K_{ij}\mc{F}_{\mc{K}},
\eeq
we can rewrite the hamiltonian $\mc{H}\equiv \mc{H}_\kappa$ as:
\beqa
&\mc{H}_\kappa \mc{F}_{\mc{K}}= \mc{H}_K \mc{F}_{\mc{K}}&\\
&\mc{H}_K =2 \left(\frac{\pi}{L}\right)^2\left[\sum_i \left(z_i \partial_i\right)^2-2\sum_{i<j}\frac{z_i z_j}{z_{ij}^2}\beta(\beta-K_{ij})\right]-E_0\, .&
\nonumber
\eeqa
This remark holds for any supersymmetric model whose hamiltonian contains a
fermionic-exchange term.

Let us denote by
$\Pi_{\mc{E}}(\mc{O})$ the projection of an operator
$\mc{O}$ on a vector space invariant under the action of
$\mc{E}$ \cite{Polychronakos:1992zk, Bernard:1993va}. For instance,
\beq
\Pi_{K}(K_{ij})=1 \mbox{ , }\qquad  \Pi_{\mc{K}}(K_{ij})=
\kappa_{ij}\mbox{ , }\qquad
\Pi_{\mc{K}}(\kappa_{ij}\kappa_{kl})=K_{kl}K_{ij} \, .
\eeq
We can consider the hamiltonian with exchange term
$K_{ij}$ as the most fundamental one.
Indeed, by an appropriate choice of
projection, various models can be generated from it. For instance,
the tCMS model and its supersymmetric generalization are obtained
respectively from:\footnote{We could also choose projections on antisymmetric
spaces,
\emph{e.g.}, $K_{ij}=-\kappa_{ij}$. This is considered in app. A.  In a similar
vein, spin degrees of freedom can be introduced in that way -- \emph{cf.} the
conclusion.}
\beqa
    \Pi_{K}(\mc{H}_K) & = & H\, ,\\
    \Pi_{\mc{K}}(\mc{H}_K) & = & \mc{H}_{\kappa}\, .
\eeqa

We now present a simple way to derive the various
types of Dunkl operators for the CMS models found in the literature
out of  the Lax operator
(\emph{cf}. \cite{Polychronakos:1992zk, Bernard:1993va, Polychronakos:1999sx}).
The `covariant' Dunkl operator is simply given by
\beq
\ba{lcl}
D_j&=&\sum_k L_{jk}(X_{jk}\rightarrow X_{jk}K_{jk})\\
&=&p_j+i\sum_{k\neq j}X_{jk}K_{jk}.
\ea
\eeq
This Dunkl operator satisfies the following properties:
\beq
\ba{cl}
    K_{ij}D_i=D_j K_{ij}& \mbox{(covariance)},\\
    \left[ D_{i}, D_{j}\right]  =-(\beta\pi/L)^2\sum_{k\neq i,j}(K_{ik}-K_{jk}) K_{ij}&\mbox{(non-commutativity)},\\
    \left[ D_i, \mc{H}_{K} \right]  = 0&\mbox{(conservation)}.
\ea
\eeq
where `covariance' means that
$D_i$ behaves like
the variable $z_i$ under the action of the symmetric group $S_N$.\footnote{Another
 simple `covariant' Dunkl operator is $\hat{D_j}=D_j \pm (\beta\pi/L)\sum_{j\neq i}K_{ij}$.
   It has the following, somewhat more natural, commutation property: $\left[ \hat{D}_{i}, \hat{D}_{j}\right]
  =\mp (2 \beta\pi/L)(\hat{D}_i-\hat{D}_j) K_{ij}$.}  It should be noted that this Dunkl operator can be viewed as the
`square' of fermionic derivatives, \emph{i.e.}, \beq
D_i=C_i^2=(C^\dagger_i)^2\, , \eeq where \beq
C_i=\frac{\partial}{\partial \theta_i}+\theta_i D_i \mtext{and}
C^\dagger_i=\theta_i+ D_i \frac{\partial}{\partial \theta_i}\,.
\eeq It is useful to introduce another Dunkl operator: \beq
\mc{D}_i=D_i\pm\frac{\pi\beta}{L}(\sum_{j<i}K_{ij}-\sum_{j>i}K_{ij}).
\eeq that satisfies \beq \ba{cl}
    K_{i,i+1}\mc{D}_{i+1}-\mc{D}_i K_{i,i+1} =\mp\frac{2\pi \beta}{L} &
\mbox{(degenerated Hecke algebra)},\\
    \left[\mc{D}_i,\mc{D}_j\right] =0&\mbox{(commutativity)},\\
    \left[\mc{D}_i,\mc{H}_{K}\right]  =  0&\mbox{(conservation)}.
\ea \eeq In addition to commuting among themselves, the $\mc{D}_i$
have the nice  property that the hamiltonian with exchange
 term $K$ lies in their universal algebra:
\beq
   \frac{1}{2} \sum_i (\mc{D}_i)^2=\mc{H}_{K}+E_0\, .
\eeq
The supersymmetric hamiltonian is recovered by a simple projection.

Using these two versions of the Dunkl operators, it is now
fairly easy to prove
the integrability of the supersymmetric trigonometric model
by constructing explicitly conserved charges from sums of powers of the
Dunkl operators.  First, the
$N$ commuting conserved bosonic quantities which generalize those of the
non-supersymmetric model
are simply\footnote{We use the same notation for the charges
constructed from the Lax operators and from the Dunkl operators. Although the lowest
order charges calculated from both expressions agree, this may not be so for the
higher-order ones. Nevertheless, they are equivalent sets of independent
charges.} :
\beqa
&\mc{H}_{(n)}=\Pi_{\mc{K}}(\sum_i\mc{D}_i^n)\mbox{ , }n=1, 2, \ldots, N&\\
&[\mc{H}_{(n)}, \mc{H}_{(m)}]=0\quad \forall n, m&\nonumber \eeqa
The proof
of the commutativity
relies on a simple property of the projections
\cite{Mathieu:2000kx}: \beq
\Pi_{\mc{K}}([A,B])=[\Pi_{\mc{K}}(A),\Pi_{\mc{K}}(B)]\qquad \mbox{ if
}\qquad [\mc{K}_{ij},A]=[\mc{K}_{ij},B]=0\, . \label{propprojec} \eeq The
operators $\mc{H}_{(n)}$ meet this requirement since $
[\mc{K}_{ij},(\sum_i\mc{D}_i^n)]=0$.
The latter property implies also $[D_i,(\sum_j\mc{D}_j^n)]=0$.  In addition to  these bosonic conserved
quantities,
charges with fermions can be constructed as:
\beq
\ba{rcl}
\mc{Q}_{(n)}&=&\Pi_{\mc{K}}(\sum_i \theta_i D_i^n) \mbox{ , }n=0,1,\ldots, N-1\, , \\
\mc{Q}_{(n)}^\dagger&=&\Pi_{\mc{K}}(\sum_i \theta_i^\dagger D_i^n) \mbox{ , }n=0,1,\ldots, N-1\, ,\\
\mc{I}_{(n)}&=&\Pi_{\mc{K}}(\sum_i
\theta_i\theta_i^\dagger D_i^n )\mbox{ , }n=0, 1, \ldots, N-1 \, .
\ea
\eeq
Note that here we use the covariant Dunkl operators rather than the
commuting ones. This is imposed by the presence of the $\theta$ factors. Indeed,
proving that those quantities are conserved still requires
 \reff{propprojec} and the covariant character of the Dunkl operators  ensures
the commutativity of the operators $\sum_i \theta_i D_i^n$ with an
arbitrary exchange operator $\mc{K}_{ij}$ (which would not be true if the
commuting Dunkl operators were used instead).

The fermionic charges are not in involution: their anticommutators
generate bosonic quantities, \emph{e.g.}, $\mc{H}_{(n)}$.  Take for instance the rational case (on the infinite line) where
 $\mc{D}_i\equiv D_i$; the conserved quantities constructed from the Dunkl operators (like those defined
from the Lax matrices) satisfy the following algebra:
\beq \label{algsusy}
\ba{c}
\{\mc{Q}_{(n)}, \mc{Q}_{(m)}^\dagger\}=
\mc{H}_{(n+m)}\, , \\
\left[\mc{Q}^\dagger_{(n)},\mc{I}_{(m)}\right]=
\mc{Q}_{(n+m)}
\mtext{,}\left[\mc{Q}_{(n)},\mc{I}_{(m)}\right]=-\mc{Q}_{(n+m)}^\dagger\, ,\\
\{\mc{Q}_{(n)}, \mc{Q}_{(m)}\}=
\{\mc{Q}_{(n)}^\dagger,\mc{Q}_{(m)}^\dagger\}=
\left[\mc{I}_{(n)},\mc{I}_{(m)}\right]=0\,
, \\
\left[\mc{Q}_{(n)},\mc{H}_{(m)}\right]=
\left[\mc{Q}^\dagger_{(n)},\mc{H}_{(m)}\right]=
\left[\mc{I}_{(n)},\mc{H}_{(m)}\right]=
\left[\mc{H}_{(n)},\mc{H}_{(m)}\right]=0\,
.
\ea \eeq
Only the last line remains true for the trigonometric and the hyperbolic models.
 In fact, it seems that the algebra of
$\{\mc{H},\mc{Q},\mc{Q}^\dagger,\mc{I}\}$ does not close linearly for those
models.

Moreover, we could also replace the $N$ independent hamiltonians
$\mc{H}_{(n)}$ by the following conserved quantities:
\beq
\mc{J}_{(n)}=\Pi_{\mc{K}}(\sum_iD_i^n )\mbox{ , }n=0, 1, \ldots, N-1 \, .
\eeq
However, the supersymmetric hamiltonian $\mc{H}$ would  not belong
to this set. There is thus some freedom in the way we choose a set
of independent conserved charges. But quite generally, the
projection $\Pi_{\mc{K}}$ of any quantity, made out of either the
$D_i$'s or the $\mc{D}_i$'s as well as out of the fermionic quantities
$\theta_i$ and $\theta_i^\dagger$, invariant under the action of
the exchange operator
 $\mc{K}_{ij}$, is always conserved in the supersymmetric model.

We end this section by mentioning that the Dunkl operators of the
transformed hamiltonian  $\bar{\mc{H}}$ are obtained as follows:
\beq \label{defDunkl} \ba{lcl}
\bar{D_i}&=&\frac{L}{2\pi}\sum_j \bar{L}_{ij}(X_{ij}\rightarrow X_{ij}K_{ij})\, ,\\
&=&z_i \partial_i+\frac{\beta}{2}\sum_{j\neq i}\frac{z_i+z_j}{z_{ij}}(1-K_{ij})\, ,\\
\bar{\mc{D}_i}&=&\bar{D_i}\pm\frac{\beta}{2}(\sum_{j<i}K_{ij}-\sum_{j>i}K_{ij})
\, ,
\ea
\eeq
and we verify that:
\beq
\Pi_{\mc{K}}(\sum_i\bar{\mc{D}_i}^2)=\bar{\mc{H}}+\frac{1}{2}\left(\frac{L}{\pi}\right)^2 E_0\, .
\eeq
All the  quantities constructed in this section
can thus be directly  transposed to the case where
the ground-state wave function is factored out.

\section{Jack superpolynomials}\label{superjack}

\subsection{Symmetry of the tCMS model's eigenfunctions }

It is known that, when defining the Jack polynomials, we
can replace condition \reff{cond1} by the condition that the
Jack polynomials be eigenfunctions of
the hamiltonian $\bar{H}$ of the
tCMS model.  Similarly, one of the conditions entering the definition
of their superanalogues will be that they be eigenfunctions of the stCMS model.
We thus begin by
making general observations regarding the symmetry properties of
the eigenfunctions of the stCMS model.

 We are looking
for  functions of $\theta$ and  $z$
that are invariant under the
transpositions
$\mc{K}_{ij}$ and that are eigenfunctions of the hamiltonian:
\beq
\bar{\mc{H}}=\sum_i (z_i \partial_i)^2+\beta \sum_{i<j}\frac{z_i+z_j}{z_{ij}}(z_i \partial_i-z_j \partial_j)-2\beta\sum_{i<j}\frac{z_i z_j}{z_{ij}^2}(1-\kappa_{ij})\, .
\eeq
Since the hamiltonian is of degree
0  in both $\theta$ and  $z$, the eigenfunctions have to be homogeneous in both
variables.
Moreover, since the
underlying mechanical
problem describes
the dynamics of
a system of particles on a circle,
the solutions must be invariant
under the transformation $x_i \to x_i+2 \pi$;  therefore, only integral
powers of the variables $z$ must be considered.
Moreover, because the product of an
 eigenfunction of degree
$n$ by a Galilean `boost',
\beq
G^q=\prod_i z_i^q,\quad q \in \mb{Z},
\eeq
gives another eigenfunction, now of degree $n+Nq$, we can restrict
 ourselves to non-negative powers of $z$.

We are
thus
seeking
polynomial eigenfunctions that are invariant under the action of
$\mc{K}_{ij}$.  This operator commutes with the superhamiltonian, which is
not the case with the operators
$K_{ij}$ and $\kappa_{ij}$ taken separately.
As already pointed out, the polynomials need to be homogeneous in $\theta$ and
$z$, let's say with degree $m$ and $n$ respectively.
These degrees
 are good quantum numbers.  Indeed,
the total `momentum'
\beq
\bar{\mc{P}}=\sum_i z_i \partial_i
\eeq
commutes with
 $\bar{\mc{H}}$ and
its eigenvalue is the degree in $z$ of the  monomial on which it acts:
\beq
\bar{\mc{P}}(z_i^{n_1}\cdots z_N^{n_N})=(\sum_i n_i)(z_i^{n_1}\cdots z_N^{n_N}).
\eeq
Likewise, the quantity
\beq
\eta=\sum_i \theta_i\theta_i^\dagger= \sum_i \theta_i{\partial
\over \partial\theta_i}
\eeq
commutes with the supersymmetric hamiltonian and counts the number of fermions
in a
monomial:
\beq
\eta (\theta_{i_1}\cdots \theta_{i_m})=m  (\theta_{i_1}\cdots \theta_{i_m}).
\eeq
We say that the above monomial belongs to the $m$-fermion sector.

We can thus solve the supersymmetric Schr\"odinger equation in
a fixed
fermionic sector at a time.  The independent eigenfunctions in a given
fermionic sector will be denoted
$\mc{A}^{(m)}_{\Lambda}(z,\theta;1/\beta)$;
they are indexed by a set of integers $\Lambda$, called a {\it
superpartition}, whose `norm' refers to the
degree in
$z$:
$|\Lambda|=n$ (see the following subsection for their actual definition).

We now clarify the symmetry properties, with respect to the $z$
variables, of any symmetric superpolynomials and, in particular, of the
eigenfunctions $\mc{A}^{(m)}_{\Lambda}$. The key observation is
that
  the solutions
 $\mc{A}^{(m)}_{\Lambda}$
must necessarily be of
 the  form:
\beq
\mc{A}^{(m)}_{\Lambda}(z,\theta;1/\beta)=\sum_{1 \leq i_1<i_2< \ldots <i_m \leq
N}\theta^{i_1 \ldots i_m}A^{i_1\ldots i_m}_{\Lambda}(z;1/\beta),
\eeq
where
\beq
\theta^{i_1\ldots i_m}=\theta_{i_1}\cdots\theta_{i_m}.
\eeq
Indeed, the various terms in the $m$-fermion sector can always be rearranged
as sums of $z$ polynomials with a monomial prefactor in the $\theta_i$'s.
$A^{i_1\ldots i_m}_{\Lambda}$ is a homogeneous polynomial in $z$ indexed by
a superpartition $\Lambda$.  The solutions $\mc{A}^{(m)}_{\Lambda}$
being symmetric
superpolynomials,
must be invariant under the action of the
 exchange operators
$\mc{K}_{ij}$.
Given that the $\theta$ products are antisymmetric, \emph{i.e.},
\beq
\kappa_{i_a i_b} \theta^{i_1\ldots i_m}=- \theta^{i_1\ldots i_m} \mbox{ if }  i_a,  i_b \in \{i_1,\ldots, i_m\}.
\eeq
the superpolynomials $ A^{i_1\ldots i_m}_{\Lambda}$ must  be
partially antisymmetric to ensure the
complete symmetry of
$\mc{A}^{(m)}_{\Lambda}$.  More precisely, the functions
$ A^{i_1\ldots i_m}_{\Lambda}$ must satisfy the following relations:
\beq
\ba{l}
K_{ij}A^{i_1\ldots i_m}_{\Lambda}(z;1/\beta)=-
A^{i_1\ldots i_m}_{\Lambda}(z;1/\beta) \quad \forall \quad i\mbox{ and }j \in
\{i_1\ldots i_m\}\, ,\\
 K_{ij}A^{i_1\ldots i_m}_{\Lambda}(z;1/\beta)=\phantom{-}
A^{i_1\ldots i_m}_{\Lambda}(z;1/\beta) \quad \forall \quad i\mbox{ and }j
\not\in \{i_1\ldots i_m\}\, .
\ea
\eeq
Note that the case
$m=1$ is special:
\beq
\ba{c}
\mc{A}^{(1)}_{\Lambda} =\sum_{i}\theta_i A^{i}_{\Lambda}(z;1/\beta)\, ,\\
K_{ij}A^{k}_{\Lambda}=A^{k}_{\Lambda}\mtext{if and only if} i,j
\neq k \, . \ea \eeq

We have thus established that any symmetric eigenfunction of the stCMS model
contains terms of mixed symmetry in $z$:  each polynomial
$ A^{i_1\ldots i_m}_{\Lambda}$ is completely antisymmetric in the variables
$\{ z_{i_1},\ldots, z_{i_m}\}$, and totally symmetric in the remaining variables
$z/\{z_{i_1},\ldots, z_{i_m}\}$.  Appendix \ref{genesjack} presents a simple way
of generating such
eigenfunctions by acting with appropriate operators on the Jack
polynomials.  However, this method does not lead to a unique characterization
of the eigenfunctions.  Later in this section will
be presented
an approach
free from this drawback.
But first, we need to define properly the superpartitions and some related
concepts.

\subsection{Symmetric superpolynomials} \label{symsuperpoly}

\noindent \emph{1- The ring of symmetric superfunctions}.  The symmetric
superpolynomials are  polynomials in
$z$ and $\theta$ that commute with the generators
 $\mc{K}_{ij}$ of the symmetric group
$\mc{S}_N$ of all possible permutations of the
$N$ variables $\zeta_i=(z_i, \theta_i)$.
As in the symmetric polynomial case, the set of all symmetric superpolynomials
in the
$N$ variables $\zeta_i$ forms a ring over the field of
integers:
\beq
\tilde{\Sigma}_N=\mb{Z}[\zeta_1, \ldots,
\zeta_N]_{\mc{S}_N}=\mb{Z}[z_1,
\ldots, z_N ; \theta_1, \ldots, \theta_N]_{\mc{S}_N}.
\eeq
It is clear that the set
of superpolynomials of degrees
$m$ in $\theta$ and $n$ in $z$ is a $\mb{Z}$-module, which we will denote
in the following way:
\beq
\tilde{\Sigma}_N^{(m;n)}=\mb{Z}[\zeta_1, \ldots, \zeta_N]_{\mc{S}_N}^{(m;n)}
\eeq
The ring of symmetric superpolynomials is thus bigraded:
\beq
\tilde{\Sigma}_N=\bigoplus_{m,n}\tilde{\Sigma}_N^{(m;n)}
\eeq

\noindent \emph{2- Superpartitions}. In the case of symmetric
polynomials, the basis elements of $\Sigma_N$ are indexed by
partitions.  In the same manner, basis elements of
$\tilde{\Sigma}_N$ can be indexed by  superpartitions. To motivate
the following definition of superpartitions, recall that the
symmetric superpolynomials in the $m$-fermion sector are
antisymmetric in the $m$ variables $\{z_{i_1},\cdots, z_{i_m}\}$
and symmetric in the remaining ones.  We thus define a
superpartition of a $m$-fermion sector as a sequence of integers
that generates two partitions separated by a semicolon: \beq
\Lambda=(\Lambda_1,\ldots,\Lambda_m;\Lambda_{m+1},\ldots,\Lambda_L)=
(\lambda^a ; \lambda^s), \eeq the first one being associated to an
antisymmetric function \beq \ba{c}
\lambda^a=(\Lambda_1,\ldots,\Lambda_m), \\
\Lambda_i>\Lambda_{i+1} \quad \forall  i=1, \ldots m-1,\\
\Lambda_i \geq 0 \quad \forall  1\leq i \leq m \, ,
\ea
\eeq
and the second one, to a symmetric function:
\beq
\ba{c}
\lambda^s= (\Lambda_{m+1},\ldots,\Lambda_L), \\
\Lambda_i \ge \Lambda_{i+1} \quad \forall i>m, \\
\Lambda_i \geq 0\mbox{ if }i=m+1\mbox{ ; } \Lambda_i > 0  \quad\forall  i> m+1.
\ea
\eeq
In the zero-fermion sector ($m=0$), the semicolon disappears and we recover the
partition $\lambda^s$.

 The length
 $L\leq N$ of a superpartition corresponds to the total number of its parts,
of which at
most two can be zero: one on the antisymmetric side and one on the
symmetric side.  The weight (or degree) of a superpartition is
simply
the sum of its parts: \beq |\Lambda|=\sum_{i=1}^L\Lambda_i
=|\lambda^a|+|\lambda^s|. \eeq For instance, the only possible
superpartitions of weight 2 in the one-fermion sector are: \beq
(2;0), \quad (0;2),\quad (1;1), \quad (0;1,1). \eeq For 2
fermions, we have instead: \beq (2,0;0) \mbox{ and } (1,0;1). \eeq
In order to specify explicitly the fermionic sector, we will
sometimes denote the degree of a superpartition as: \beq
\mbox{degree } ( m;n ) \Leftrightarrow (\mbox{fermionic sector } m
; \mbox{weight }n=|\Lambda|) \, .\eeq
Summation formulas giving
the number of superpartitions of degree $(m;n)$ are
presented in appendix \ref{compteetats}.

We mention finally that
to any  superpartition $\Lambda$ there corresponds a single standard
partition $\lambda$ obtained by rearranging the parts of the
superpartition in decreasing order:
\beq
\lambda=\{\lambda_i | \lambda_i \in \{\Lambda_1, \ldots, \Lambda_L\}, \lambda_i \geq \lambda_{i+1}\}\, .
\eeq

\noindent \emph{3- Monomial symmetric superpolynomials}. We can now introduce a
basis of
$\tilde{\Sigma}_N^{(m;n)}$
that generalizes the symmetric monomial basis of $\Sigma_N^{(n)}$: \beq
m_{\Lambda}^{(m)}(z,\theta)=\sm{\Lambda_1,\ldots,
\Lambda_m;\Lambda_{m+1},\ldots,\Lambda_L}(z,\theta)={\sum_{P\in
S_N}}'\theta^{P(1, \ldots, m)}z^{P(\Lambda)}, \eeq  (recall that the prime
indicates that the  summation is
restricted to distinct terms). It is understood that the action of the
permutations on a superpartition is not affected by the semicolon: \beq
P(\Lambda)=(\Lambda_{P(1)},\ldots,\Lambda_{P(m)};\Lambda_{P(m+1)},\ldots,\Lambda_{P(L)}).
\eeq The functions $m_{\Lambda}^{(m)}$ are called \emph{monomial
symmetric superfunctions} and $m_{\Lambda}^{(0)} \equiv m_\lambda(z) $. More explicitly, the
monomial superfunctions can be written in the following way: \beqa
\sm{\Lambda_1,\ldots,\Lambda_m;\Lambda_{m+1},\ldots,\Lambda_L} & =
&
\sm{\lambda^a;\lambda^s}\\
&=&\sum_{ i_1<i_2<\ldots <i_m}\theta^{i_1,\ldots,i_m}
a_{\lambda^a}(z_{i_1},\ldots, z_{i_m})
m_{\lambda^s}(z/\{z_{i_1},\ldots, z_{i_m}\}) \, ,\nonumber\\
&=&\sum_{ i_1<\ldots <i_m}\sum_{P^a\in S_m}\sum_{P^s \in
S_{N-m}}\mbox{sgn}(P^a) \theta_{i_1}\ldots \theta_{i_m}
z_{i_1}^{\Lambda_{P^a(1)}}\cdots z_{i_m}^{ \Lambda_{P^a(m)}}
z_{i_m+1}^{ \Lambda_{P^s(m+1)}}\cdots z_{i_L}^{\Lambda_{P^s(L)}}\,
,\nonumber \eeqa where we have introduced the antisymmetric
monomial function \beq \label{antisimf} a_{\lambda}
(z_1,\ldots,z_N) =  {\sum_{P\in
S_N}}'\mbox{sgn}(P)z^{P(\lambda)}={\sum_{P\in S_N}}'\mbox{sgn}(P)
z_1^{\lambda_{P(1)}}\cdots z_N^{\lambda_{P(N)}} \, . \eeq Many
examples of monomial superfunctions are shown in Table
\ref{tabsmono}.

\renewcommand{\baselinestretch}{1.0}
\small\normalsize

\bt
\caption{List of all the monomial superfunctions of weight $|\Lambda|\leq 3$
for  $N={\rm max}\, (3,L)$  variables}
\small
\bc
\label{tabsmono}
\btab{|c||c|c|c|}
\hline
Weight & Sector& Superpartition &Monomial superfunction \\
$|\Lambda|$& $m$ &$\Lambda$&$m^{(m)}_{\Lambda}(z,\theta)$\\
\hline\hline
0 & 1 & (0;0) &$ \theta_1+\theta_2+\theta_3 $\\
 \hline \hline
 1& 1 & (1;0) & $ \theta_1 z_1+\theta_2 z_2+ \theta_3 z_3 $ \\
 &    & (0;1) & $ \theta_1(z_2+z_3)+\theta_2(z_1+z_3)+\theta_3(z_1+z_2) $\\
 & 2 & (1,0;0) & $ \theta_1 \theta_2 (z_1-z_2) +\theta_1 \theta_3 (z_1-z_3) +\theta_2 \theta_3 (z_2-z_3) $ \\
\hline \hline
2 & 1 & (1;1)& $ \theta_1z_1(z_2+z_3)+\theta_2 z_2(z_1+z_3)+\theta_3 z_3(z_1+z_2)$ \\
 &  & (0;1,1) &$ \theta_1(z_2 z_3)+\theta_2 (z_1 z_3) \theta_3(z_1 z_2) $ \\
 & 2 & (1,0;1) & $ \theta_1 \theta_2 (z_1-z_2)z_3 +\theta_1 \theta_3 (z_1-z_3) z_2 +\theta_2\theta_3 (z_2-z_3) z_1 $ \\
\hline
 & 1 & (2;0) & $ \theta_1 z_1^2 + \theta_2 z_2^2 + \theta_3 z_3^2 $\\
 &   & (0;2) & $ \theta_1(z_2^2+z_3^2)+\theta_2(z_1^2+z_3^2)+\theta_3(z_1^2+z_2^2)$\\
 & 2 & (2,0;0) & $ \theta_1 \theta_2 (z_1^2-z_2^2) +\theta_1 \theta_3 (z_1^2-z_3^2) +\theta_2 \theta_3 (z_2^2-z_3^2) $ \\
\hline \hline
3 & 1 & (1;1,1) & $ \theta_1 z_1(z_2 z_3+z_2 z_4+z_3 z_4)+ \theta_2 z_2(z_1 z_3+z_1 z_4+z_3 z_4) $\\ & & &$+ \theta_3 z_3(z_1 z_2+z_1 z_4+z_2 z_4) +\theta_4 z_4 (z_1 z_2+z_1 z_3+z_2 z_3)$ \\
 &   & (0;1,1,1) & $ \theta_1 (z_2 z_3 z_4)+ \theta_2 (z_1 z_3 z_4)+ \theta_3 (z_1 z_2 z_4)+\theta_4(z_1 z_2 z_3) $ \\
 & 2 & (1,0;1,1) & $ \theta_1 \theta_2 (z_1-z_2)(z_3 z_4)+\theta_1 \theta_3 (z_1-z_3)(z_2 z_4) $ \\ & & & $+\theta_1 \theta_4 (z_1-z_4)(z_2 z_3)+\theta_2 \theta_3 (z_2-z_3)(z_1 z_4) $ \\ & & &$+\theta_2 \theta_4 (z_2-z_4)(z_1 z_3)+\theta_3 \theta_4 (z_3-z_4)(z_1 z_2) $ \\
\hline
 & 1  & (2;1) & $ \theta_1 z_1^2 (z_2+z_3)+ \theta_2 z_2^2 (z_1+z_3)+ \theta_3 z_3^2 (z_1+z_2) $ \\
&   & (1;2) & $ \theta_1 z_1 (z_2^2+z_3^2)+ \theta_2 z_2(z_1^2+z_3^2)+ \theta_3 z_3 (z_1^2+z_2^2) $ \\
&   & (0;2,1) & $ \theta_1 (z_2^2 z_3+z_2 z_3^2)+ \theta_2 (z_1^2 z_3+z_1 z_3^2)+\theta_3 (z_1^2 z_2+z_1 z_2^2) $ \\
 & 2 & (2,1;0) & $ \theta_1 \theta_2 (z_1^2 z_2 -z_1 z_2^2)+\theta_1 \theta_3 (z_1^2 z_3 -z_1 z_3^2)+\theta_2 \theta_3 (z_2^2 z_3 -z_2 z_3^2)$ \\
 &  & (2,0;1) & $ \theta_1 \theta_2 (z_1^2  - z_2^2)(z_3)+ \theta_1 \theta_3 (z_1^2  - z_3^2)(z_2)+ \theta_2 \theta_3 (z_2^2  - z_3^2)(z_1)$ \\
 &  & (1,0;2) & $ \theta_1 \theta_2 (z_1 - z_2)(z_3^2)+ \theta_1 \theta_3 (z_1- z_3)(z_2^2)+ \theta_2 \theta_3 (z_2 - z_3)(z_1^2)$ \\
\hline
 & 1  & (3;0) & $ \theta_1 z_1^3+ \theta_2 z_2^3 + \theta_3 z_3^3 $ \\
 &   & (0;3) & $ \theta_1 (z_2^3+z_3^3)+ \theta_2 (z_1^3+z_3^3)+ \theta_3 (z_1^3+z_2^3) $ \\
 & 2 & (3,0;0) & $ \theta_1 \theta_2 (z_1^3-z_2^3)+ \theta_1 \theta_3 (z_1^3-z_3^3)+ \theta_2 \theta_3 (z_2^3-z_3^3) $ \\
\hline
\etab
\ec
\et
\renewcommand{\baselinestretch}{1.0}
\small\normalsize

\subsection{Jack superpolynomials: monomial expansion}\label{sjackmon}

We now define the {\it Jack superpolynomials} in the $m$-fermion sector as the
unique eigenfunctions of the supersymmetric hamiltonian
 $\bar{\mc{H}}$
 that can be decomposed in terms of monomial
superfunctions in the following way: \beq \label{sjackensm}
\mc{J}_\Lambda^{(m)}(z,\theta;1/\beta)=m_{\Lambda}^{(m)}(z,\theta)+\sum_{\omega
< \lambda}c_{\Lambda,\Omega}(\beta) m_{\Omega}^{(m)}(z,\theta)\, ,
\eeq where $\omega$ and $\lambda$ are the partitions associated to
the rearrangements of $\Omega$ and $\lambda$ respectively
\footnote{The reader is referred to \cite{DLM} for a proof that
this definition does in fact characterize a family of polynomials
that forms a basis of $\tilde{\Sigma}_N^{(m;n)}$.}.

The relation between Jack superpolynomials and the
usual Jack polynomials can now be stated precisely:
$\mc{J}^{(0)}_{\Lambda}=J_{\lambda^s}$.

 The coefficients
$c_{\Lambda,\Omega}(\beta)$  in \reff{sjackensm} are rational functions
in $\beta$. We can
easily verify, from the leading terms, that the spectrum of the supersymmetric hamiltonian
 $\bar{ \mc{H}}$ is the same as that of
$\bar{H}$, \emph{i.e.},
\beq
\bar{\mc{H}}\mc{J}_\Lambda^{(m)}(z,\theta;1/\beta)=
\varepsilon_\Lambda\mc{J}_\Lambda^{(m)}(z,\theta;1/\beta)=
\varepsilon_\lambda\mc{J}_\Lambda^{(m)}(z,\theta;1/\beta)\,,
 \eeq
where the eigenvalues are given by: \beq
\varepsilon_\lambda=\sum_j [\lambda_j^2+\beta(N+1-2j)\lambda_j],
\eeq
$\lambda$ being the rearrangement of $\Lambda$. Observe
that the eigenvalue $\varepsilon_\lambda$ is independent of the
fermionic sector, \emph{i.e.}, independent of
the value of $m$.\footnote{
Therefore, the most general eigenfunction having energy $\varepsilon_{\Lambda}$
is a linear combination of all the eigenfunctions
$\mc{J}^{(m)}_{\Lambda}$ whose eigenvalue is
$\varepsilon_{\Lambda}=\varepsilon_{\lambda}$ :
\beq\nonumber
\ba{lcl}
\mc{J}_{\lambda}(z,\theta;1/\beta) &=&\sum_{m=0}^N \sum_{\Lambda}
\tau^m_{\Lambda}
\mc{J}^{(m)}_{\Lambda}(z,\theta;1/\beta) \label{staylor}\\
&=& \jack+\sum_{\Lambda} [\tau^1_ {\Lambda}\theta^i
J^{i}_{\Lambda}(z;1/\beta)+\tau^2_ {\Lambda}\theta^{i,j}
J^{i,j}_{\Lambda}(z;1/\beta)+\ldots+\tau^N_
{\Lambda}\theta^{1\ldots N} J^{1 \ldots
N}_{\Lambda}(z;1/\beta)]
\ea
\eeq
 where an ordered summation
on repeated indices is understood:
\beq
\nonumber \theta^{i_1\ldots
i_m}J^{i_1 \ldots i_m}_{\Lambda} =\sum_{1 \leq i_1<i_2< \ldots
<i_m \leq N}\theta_{i_1} \cdots \theta_{i_m}J^{i_1 \ldots
i_m}_{\Lambda} \, .
 \eeq
Here $\tau^m_{\Lambda}$ stands for a
commuting (anticommuting) constant when $m$ is even (odd).  These constants are
auxiliary: they only guarantee the homogeneity of the statistics of the
superpolynomials. Equation  \reff{staylor} thus corresponds in
some way to the Taylor series of a generalized Jack superpolynomial
around $\theta=0$.  The first term in the expansion is simply a
Jack polynomial.}

Tables \ref{tabsjack1} and \ref{tabsjack2} present simple examples
whose degrees $(m;n)$ are not larger than $(3;4)$. The
coefficients $c_{\Lambda,\Omega}$ are obtained by simply
diagonalizing the hamiltonian. Polynomials with the same partition
have the same eigenvalue.  Given that the eigenvalues are
independent of $m$, they can be read in Table \ref{tabjacks}.

 We should stress that the decomposition is not simply a
 straightforward extension of the triangular decomposition of the
 Jack polynomials, where the ordering is on partitions.
Here the `ordering' that allows a triangular decomposition is on
superpartitions, albeit rearranged.  The existence of such an
`ordering' seems to us  quite remarkable (even though it is not a
genuine ordering, as it is shown in \cite{DLM}).

A closer look at those results shows that
expression
 \reff{sjackensm} is not restrictive enough: certain monomials allowed by the
dominance ordering of the rearranged superpartitions do not appear in the actual
expansion of the Jack superpolynomials.  For instance, no monomial superfunction
associated to a superpartition with a $0$ on the antisymmetric side appears
in the expansion of a Jack superpolynomial whose superpartition does not
contain any $0$ to the left of the semicolon. This information is not
however encoded in eq. \reff{sjackensm}.  A partial ordering formulated directly
among superpartitions would lead to a more precise formulation of the monomial
expansion of the Jack superpolynomials. Such an ordering has indeed been found
\cite{DLM}. But because its formulation is somewhat technical, it will be
presented elsewhere.

\renewcommand{\baselinestretch}{1.0}
\small\normalsize
\bt
\caption{The Jack superpolynomials of weight $|\Lambda|\leq 3$}
\bc
\label{tabsjack1}
\btab{|c||c|c|c|c|}
\hline
Weight & Partition & Sector& Superpartition &Jack superpolynomial  \\
$|\Lambda|$&$\lambda$& $m$ &$\Lambda$&$\mc{J}^{(m)}_{\Lambda}(z,\theta;1/\beta)$\\
\hline\hline
0 &$(0)$& 1 &$ (0;0)$ &$\sm{0;0} $\\
 \hline \hline
1 & $(1)$& 1 & $(1;0) $& $ \sm{1;0}$ \\
 & &   & $(0;1)$ & $ \sm{0;1}$\\
 & &2 & $(1,0;0) $& $ \sm{1,0;0}$ \\
\hline \hline
2 & $(1^2)$&1 & $(1;1)$& $ \sm{1;1}$ \\
 &  && $(0;1^2)$ &$ \sm{0;1^2}$ \\
 & &2 & $(1,0;1)$ & $ \sm{1,0;1}$ \\
\hline
 &$(2)$& 1 & $(2;0) $& $ \sm{2;0}+\frac{\beta}{1+\beta}\sm{1;1} $\\
 &   & & $(0;2)$ & $ \sm{0;2}+\frac{\beta}{1+\beta}\sm{0;1^2}+\frac{2\beta}{1+\beta}\sm{0;1^2} $\\
 & & 2 & $(2,0;0)$ & $ \sm{2,0;0}+\frac{\beta}{1+\beta}\sm{1,0;1}$ \\
\hline \hline
3 &$(1^3)$& 1 & $(1;1^2)$ & $ \sm{1;1^2}$\\
 &  & & $(0;1^3) $& $ \sm{0;1^3} $ \\
 & &2 & $(1,0;1^2)$ & $ \sm{1,0;1^2}$ \\
\hline
 & $(2,1)$&1  & $(2;1) $& $  m_{(2;1)}+\frac{2 \beta}{1+2\beta}m_{(1;1^2)}$ \\
&  & & $(1;2)$ & $  m_{(1;2)}+\frac{2\beta}{1+2\beta}m_{(1;1^2)}$ \\
&  & & $(0;2,1) $& $  m_{(0;2,1)}+\frac{2\beta}{1+2\beta}m_{(1;1^2)}+\frac{6\beta}{1+2\beta}m_{(0;1^3)}$ \\
 & &2 & $(2,1;0) $& $ \sm{2,1;0}$ \\
 &  & & $(2,0;1) $& $\sm{2,0;1}+\frac{2 \beta}{1+2\beta}\sm{1,0;1^2} $ \\
 &  & & $(1,0;2)$ & $\sm{1,0;2}+\frac{2 \beta}{1+2\beta}\sm{1,0;1^2}  $ \\
\hline
 &$(3)$& 1  & $(3;0)$ & $\sm{3;0}+\frac{2\beta}{2+\beta} \sm{2;1}+\frac{\beta}{2+\beta} \sm{1;2}+\frac{2\beta^2}{(1+\beta)(2+\beta)} \sm{1;1^2}$ \\
 & &  & $(0;3)$ & $\sm{0;3}+\frac{\beta}{2+\beta} \sm{2;1}+\frac{3\beta}{2+\beta} \sm{0;2,1}+\frac{2\beta}{2+\beta} \sm{1;2}$\\
& & & &$+\frac{4\beta^2}{(1+\beta)(2+\beta)} \sm{1;1^2}+\frac{6\beta^2}{(1+\beta)(2+\beta)} \sm{0;1^3}$ \\
 & &2 & $(3,0;0)$ & $ \sm{3,0;0}+\frac{\beta}{2+\beta}\sm{2,1;0}+\frac{2\beta}{2+\beta}\sm{2,0;1}+\frac{\beta}{2+\beta}\sm{1,0;2}$\\
& & & & $+\frac{2\beta^2}{(1+\beta)(2+\beta)}\sm{1,0;1^2}$ \\
\hline
\etab
\ec
\et
\renewcommand{\baselinestretch}{1.0}
\small\normalsize

\renewcommand{\baselinestretch}{1.0}
\small\normalsize

\bt
\caption{Jack superpolynomials of weight $|\Lambda|=4$}
\bc
\label{tabsjack2}
\btab{|c|c|c|c|}
\hline
 Partition & Sector& Superpartition &Jack superpolynomial \\
$\lambda$& $m$ &$\Lambda$&$\mc{J}^{(m)}_{\Lambda}(z,\theta;1/\beta)$\\
\hline\hline

 $(1^4)$& 1 &$ (1;1^3)$ &$\sm{1;1^3} $\\
   & & $(0;1^4) $&$\sm{0;1^4} $\\
 &2 & $(1,0;1^3)$ &$\sm{1,0;1^3} $\\
\hline
$(2,1^2)$ & 1 & $(2;1^2) $&$
m_{(2;1^2)}+\frac{3\beta}{1+3\beta}\sm{1;1^3} $ \\
  & & $(1;2,1) $&$
m_{(1;2,1)}+\frac{6\beta}{1+3\beta}\sm{1;1^3} $ \\
 & & $(0;2,1^2)$ &$
m_{(0;2,1^2)}+\frac{3\beta}{1+3\beta}\sm{1;1^3}+\frac{12\beta}{1+3\beta}\sm{0;1^4} $ \\
& 2 & $(2,1;1)$ &$ \sm{2,1;1} $ \\
 &  & $(2,0;1^2) $&$\sm{2,0;1^2}+\frac{3 \beta}{1+3\beta}\sm{1,0;1^3} $ \\
  & & $(1,0;2,1)$ &$\sm{1,0;2,1}+\frac{6 \beta}{1+3\beta}\sm{1,0;1^3} $ \\
& 3 & $(2,1,0;1)$ & $\sm{2,1,0;1}$\\

\hline
$(2^2)$ & 1 &$(2;2)$ &  $\sm{2;2}+\frac{2\beta}{1+\beta}\sm{2;1^2}+\frac{\beta}{1+\beta}\sm{1;2,1}+\frac{6\beta^2}{(1+\beta)(1+2\beta)}\sm{1;1^3} $ \\
  &&$(0;2^2)$ &  $\sm{0;2,2}+\frac{\beta}{1+\beta}\sm{1;2,1}+\frac{2\beta}{1+\beta}\sm{0;2,1^2}+\frac{12\beta^2}{(1+\beta)(1+2\beta)}\sm{0;1^4} $\\
& & &$+\frac{6\beta^2}{(1+\beta)(1+2\beta)}\sm{1;1^3}$ \\
&2 &$(2,0;2)$ &  $\sm{2,0;2}+\frac{\beta}{1+\beta}\sm{2,1;1}+\frac{2\beta}{1+\beta}\sm{2,0;1^2}+\frac{\beta}{1+\beta}\sm{1,0;2,1}$\\
& & &$+\frac{6\beta^2}{(1+\beta)(1+2\beta)}\sm{1,0;1^3} $ \\

\hline
 $(31)$ &1  & $(3;1)$ &  $\sm{3;1}+\frac{\beta}{1+\beta}\sm{2;2}+\frac{\beta(2+3\beta)}{(1+\beta)^2}\sm{2;1^2}+\frac{\beta(1+2\beta)}{2(1+\beta)^2}\sm{1;2,1}$\\
&& &$+\frac{3\beta^2}{(1+\beta)^2}\sm{1;1^3} $ \\
  &  & $(1;3)$ &  $\sm{1;3}+\frac{\beta}{1+\beta}\sm{2;2}+\frac{\beta^2}{(1+\beta)^2}\sm{2;1^2}+\frac{\beta(3+4\beta)}{2(1+\beta)^2}\sm{1;2,1}$\\
& &&$+\frac{3\beta^2}{(1+\beta)^2}\sm{1;1^3} $ \\
   && $(0;3,1)$ &  $\sm{0;3,1}+\frac{2\beta}{1+\beta}\sm{0;2,2}+\frac{\beta}{(1+\beta)}\sm{2;1^2}+\frac{\beta(1+2\beta)}{(1+\beta)^2}\sm{1;2,1}$\\
& & &$+\frac{\beta(3+5\beta)}{(1+\beta)^2}\sm{0;2,1^2}+\frac{6\beta^2}{(1+\beta)^2}\sm{1;1^3} +\frac{12\beta^2}{(1+\beta)^2}\sm{1;1^3} $ \\

\hline
&2&$ (3,1;0)$ & $\sm{3,1;0}+\frac{\beta}{1+\beta}\sm{2,1;1}$\\
  & &$(3,0;1) $& $\sm{3,0;1}+\frac{\beta}{1+\beta}\sm{2,0;2}+\frac{\beta(1+2\beta)}{2(1+\beta)^2}\sm{2,1;1}+\frac{\beta(2+3\beta)}{(1+\beta)^2}\sm{2,0;1^2}$\\
& & &$+\frac{\beta(1+2\beta)}{2(1+\beta)^2}\sm{1,0;2,1}+\frac{3\beta^2}{(1+\beta)^2}\sm{1,0;1^3}$\\
   && (1,0;3) & $\sm{1,0;3}+\frac{\beta}{1+\beta}\sm{2,0;2}-\frac{\beta}{2(1+\beta)^2}\sm{2,1;1}+\frac{\beta^2}{(1+\beta)^2}\sm{2,0;1^2}$\\
 & && $+\frac{\beta(3+4\beta)}{2(1+\beta)^2}\sm{1,0;2,1}+\frac{3\beta^2}{(1+\beta)^2}\sm{1,0;1^3}$\\

&3 & $(3,1,0;0)$ & $\sm{3,1,0;0}+\frac{\beta}{1+\beta}\sm{2,1,0;1}$\\

\hline
 $(4)$ &  1 &$(4;0)$ &$m_{(4;0)} +{\frac{3 \beta  }{3 + \beta }} m_{(3;1)}+{\frac{\beta  }{3 + \beta }} m_{(1;3)}+
  {\frac{3 \beta  \left( 1 + \beta  \right)  }
    {\left( 2 + \beta  \right)  \left( 3 + \beta  \right) }
    } m_{(2;2)}$\\
&&&$+
  {\frac{3 {{\beta }^2} }
    {\left( 2 + \beta  \right)  \left( 3 + \beta  \right) }
    } m_{(1;2,1)}+ {\frac{6 {{\beta }^2} }
    {\left( 2 + \beta  \right)  \left( 3 + \beta  \right) }
    } m_{(2;1^2)}+ {\frac{6 {{\beta }^3} }
    {\left( 1 + \beta  \right)
      \left( 2 + \beta  \right)  \left( 3 + \beta  \right)
      }}m_{(1;1^3)}$
\\
  & &$(0;4)$ & $m_{(0;4)} +{\frac{\beta  }{3 + \beta }} m_{(3;1)}+  {\frac{3 \beta }{3 + \beta }}  m_{(1;3)}$\\
&&&$+{\frac{4 \beta }{3 + \beta }} m_{(0;3,1)} +
  {\frac{3 \beta  \left( 1 + \beta  \right)  }
    {\left( 2 + \beta  \right)  \left( 3 + \beta  \right) }
    } m_{(2;2)}+ {\frac{6 \beta  \left( 1 + \beta  \right)
     }{\left( 2 + \beta  \right)
      \left( 3 + \beta  \right) }}  m_{(0;2,2)}$\\
&&&$+
  {\frac{9 {{\beta }^2}}
    {\left( 2 + \beta  \right)  \left( 3 + \beta  \right) }
    } m_{(1;2,1)} + {\frac{6 {{\beta }^2}}
    {\left( 2 + \beta  \right)  \left( 3 + \beta  \right) }
    }  m_{(2;1^2)}+ {\frac{12 {{\beta }^2} }
    {\left( 2 + \beta  \right)  \left( 3 + \beta  \right) }
    } m_{(0;2,1^2)}$\\
&&&$+
  {\frac{18 {{\beta }^3} }
    {\left( 1 + \beta  \right)
      \left( 2 + \beta  \right)  \left( 3 + \beta  \right)
      }}m_{(1;1^3)} + {\frac{24 {{\beta }^3} }
    {\left( 1 + \beta  \right)
      \left( 2 + \beta  \right)  \left( 3 + \beta  \right)
      }}m_{(0;1^4)}
 $ \\

&2 & $(4,0;0)$ &$\sm{4,0;0}+\frac{2\beta}{3+\beta}\sm{3,1;0}+\frac{3\beta}{3+\beta}\sm{3,0;1}+\frac{\beta}{3+\beta}\sm{1,0;3}$\\
&& &
$+\frac{3\beta(1+\beta)}{(2+\beta)(3+\beta)}\sm{2,0;2}+\frac{3\beta^2}{(2+\beta)(3+\beta)}\sm{2,1;1}+\frac{6\beta^2}{(2+\beta)(3+\beta)}\sm{2,0;1^2}$\\
& &&
$+\frac{3\beta^2}{(2+\beta)(3+\beta)}\sm{1,0;2,1}+\frac{6\beta^3}{(1+\beta)(2+\beta)(3+\beta)}\sm{1,0;1^3} $\\
\hline
\etab
\ec
\et

\renewcommand{\baselinestretch}{1.0}
\small\normalsize

\section{Conclusion}

In this work, we have presented a number of results concerning the stCMS
model: its reformulation in terms of the exchange-operator formalism,  the Lax
formalism, the Dunkl operators and an explicit construction for the conserved
charges. In fact, $4N$ conserved charges have been constructed, $2N$ bosonic and
$2N$ fermionic ones.

However, our most important results pertain to the construction of the stCMS
eigenfunctions, with particular emphasis on the subclass  which  we call the
Jack superpolynomials and which is a  natural generalization of the Jack
polynomials. In view of defining them properly, we have introduced the pivotal
concept of superpartitions providing the natural labelling of the Jack
superpolynomials. The Jack superpolynomials are then naturally defined
by further imposing that they decompose in a specific manner in terms of monomial
superfunctions, a procedure that extends  the standard way of
defining the Jack polynomials.

In a forthcoming publication, we will present a dominance ordering
on the superpartitions that suggests an exact expression
for the coefficients $c_{\Omega,\Lambda}$ in \reff{sjackensm} and a related
determinantal formula.  These results can in turn be used to
demonstrate the actual existence of the Jack superpolynomials.

Numerous extensions of this work can be contemplated.  The most
immediate one concerns the generalization from $2$ to an arbitrary
even number $2\mc{M}$ of supersymmetries. This is rather
straightforward in the exchange-operator formalism:   it suffices
to set \beqa
\mc{K}_{ij}&\equiv&\kappa_{ij}K_{ij}\, , \\
\kappa_{ij}&\equiv&\kappa^1_{ij} \ldots \kappa^{\mc{M}}_{ij}\, , \eeqa
where $\kappa^a_{ij}$ is the operator that exchanges the
Grassmannian variables $\theta^a$ and $\theta^{a\dagger}$, where $a=1,\cdots,
\mc{M}$.   The
$2\mc{M}$ generators of the supersymmetric transformations are
then: \beqa
&\mc{Q}^a=\Pi_{\mc{K}}(\sum_i \theta^a_i D_i)\, ,&\\
&\mc{Q}^{a\dagger}=\Pi_{\mc{K}}(\sum_i \theta_i^{a\dagger} D_i)\, .&
\eeqa  The construction of the conserved quantities
is analogous to the one we have presented in the case with
$2\mc{M}=2$ supersymmetries.
 The construction of the eigenfunctions is also rather direct: the
Jack superpolynomials are then indexed by $\mc{M}$ fermionic
sectors and a superpartition involving $\mc{M}$ antisymmetric
partitions: \beqa
\mc{J}^{(m)}_{\Lambda}&\rightarrow& \mc{J}^{(m_1, \ldots,
m_{\mc{M}})}_{\Lambda}\, ,\\ &\rightarrow& \mc{J}_{(\Lambda_1, \ldots,
\Lambda_{m_1};\Lambda_{m_1+1},\ldots,
\Lambda_{m_1+m_2};\ldots;\Lambda_{m_1+\ldots+m_{\mc{M}}+1}, \ldots,\Lambda_L)}
\, . \nonumber
\eeqa

Another simple way of extending the model is by adding spin
degrees of freedom. Again, this is very simple in the framework of
the exchange-operator formalism, where we only need to  add an
extra piece in the total exchange operator, \emph{i.e.}, set: \beq
\mc{K}_{ij}=\kappa_{ij}K_{ij}\sigma_{ij}\, . \eeq $\sigma_{ij}$
interchanges the spins of particles $i$ and $j$. This leads to a
supersymmetric model with spin degrees of freedom. The Lax
formalism and the construction of the Dunkl operators can be
extended directly to this more general case. Note that  the
spectrum of a supersymmetric model, with or without spin, is
always the same as its non-supersymmetric relative. The effect of
supersymmetry on the spectrum is to increase the degeneracy of
each eigenvalue, in addition to generate fermionic states.  But
this does not change the fractional statistics,
 \emph{i.e.}, the existence of a generalized Pauli principle.

Another natural generalization concerns the application of the
method developed here, to the construction of the eigenfunctions
of the supersymmetric rational CMS model with harmonic term,
thereby generating the Hi-Jack (or generalized Hermite)
superpolynomials. Similarly, one could work out the super
extension of the Macdonald polynomials as the eigenfunctions of
the supersymmetric Ruijsenaars-Schneider model.

Coming back to the Jack superpolynomials \emph{per se}, we can also point out
various axes of research stemming from this work. At first, it would be
interesting to reformulate their definition in a more abstract way. Mimicking the
mathematical definition of the Jack polynomials, this would amount to define the
superpolynomials in terms of two requirements: triangularity and
orthogonality. A natural way of defining a scalar product is not difficult to
figure out.
However, one would need to lift the degeneracy of the Jack superpolynomials. A
natural idea for this would be to look for combinations of the Jack
superpolynomials that are eigenfunctions of the charges $\mc{I}_n$.

Another interesting issue would be to try to formulate a sort of superfield
formalism, by  summing  the Jack polynomials associated to a given
superpartition over the different fermionic
sectors.  This would construct what we could call a super-Jack polynomial.
At this point, this summation over
the different sectors appears to be loosely defined, however. How could we
constraint the value of the relative  coefficients associated to the different
fermionic sectors?

We also expect that there exists creation operators that provide the super
analogues of the operators constructed in \cite{Lapointe1995}. A first trial in
that direction has been presented in app. B. This indeed maps simple stCMS
eigenfunctions -- in fact, Jack polynomials, hence Jack superpolynomials
specialized to the zero-fermion sector -- to other stCMS eigenfunctions.
However, as we have already pointed out, the action of these creation operators
does not close within the set of Jack superpolynomials labelled by
superpartitions.

Two other issues, both rooted in our initial motivations for
studying sCMS models, are worth mentioning. Given the rather
intriguing relation that exists between the Virasoro singular
vectors and the Jack polynomials, one can naturally ask whether
there is a similar relation at the supersymmetric level. But this
would presumably require the definition of a proper super-Jack
polynomial (\emph{i.e.}, an appropriate sum over the different fermionic
sectors of the Jack superpolynomials). Note that the potentially
related singular vectors would be those of the $\mc{N}=2$
superconformal algebra.

On the other hand, we have recalled in the introduction the
remarkable connection relating the dynamics of
the rational CMS models
and the time evolution of the rational solutions of the Korteweg-de Vries equation.
A quite  natural quest would be to try to find a supersymmetric counterpart to this
phenomenon.

We expect to report elsewhere on some of these other issues.

\begin{appendix}

\section{Antisymmetric eigenfunctions} \label{etatsantisym}

In this appendix, we construct excited states
 related to the second ground state
$\tilde{\psi}_0=\Delta^{-\beta}\ket{\tilde{0}}$.
The differential
representation $\ket{\tilde{0}}\rightarrow \theta_1\cdots\theta_N$ makes clear
the complete antisymmetry of the fermionic ground state $\ket{\tilde{0}}$.
Thus, the exited states that behave like $\tilde{\psi}_0$ under the exchange of
bosonic and fermionic variables take the following form:
\beq
\tilde{\psi}(z,\theta)=\phi (z,\theta^\dagger)
\tilde{\psi}_0=\phi (z,\partial/\partial
\theta)\Delta^{-\beta}\theta_1\cdots\theta_N \mtext{where} \mc{K}_{ij}\phi =\phi
\mc{K}_{ij}.
\eeq
Equivalently, we can define $\tilde{\psi}_\Lambda$ as:
\beq
\tilde{\psi} (z,\theta)=\tilde{\phi}(z,\theta)
\Delta^{-\beta}\mtext{where} \mc{K}_{ij}\tilde{\phi}  =-\tilde{\phi} \mc{K}_{ij}.
\eeq

The antisymmetric states $\tilde{\phi}$ are eigenfunctions of the
supersymmetric hamiltonian:
\beq
\ba{rcl}
\tilde{\mc{H}}&=&\frac{1}{2}\left(\frac{L}{\pi}\right)^2
\Delta^{\beta}\mc{H}\Delta^{-\beta},\\
&=&\sum_i (z_i \partial_i)^2-\beta \sum_{i<j}\frac{z_i+z_j}{z_{ij}}
(z_i \partial_i-z_j\partial_j)+2\beta\sum_{i<j}\frac{z_i
z_j}{z_{ij}^2}(1+\kappa_{ij}),\\
&=&\bar{\mc{H}}(\beta\rightarrow-\beta,\kappa_{ij}\rightarrow-\kappa_{ij})\, .
\ea
\eeq
Since $\kappa_{ij}\tilde{\phi}=-K_{ij}\tilde{\phi}$, the superfunctions
are also eigenfunctions of the (modified) hamiltonian with an exchange term:
\beq
\tilde{\mc{H}}\tilde{\phi}=\bar{\mc{H}}^\prime_K\tilde{\phi},
\eeq
where the prime symbol indicates that the sign of the coupling constant $\beta$
is reversed:
$f^\prime(\beta)=f(-\beta)$.

Like the symmetric superfunctions, the  antisymmetric superfunctions
$\tilde{\phi}$ of degree $(m;n)$ can be indexed by a
superpartition $\Lambda$ and denoted
$\tilde{\mc{J}}_\Lambda^{(m)}$.  These solutions will be called the
{\it  antisymmetric Jack superpolynomials}.  They can be defined
in terms of a triangular decomposition in antisymmetric monomial superfunctions:
\beq
\tilde{\mc{J}}_\Lambda^{(m)}(z,\theta;1/\beta)=a_{\Lambda}^{(m)}(z,\theta)+\sum_{\omega
< \lambda}c_{\Lambda,\Omega}(\beta) a_{\Omega}^{(m)}(z,\theta)\, ,
\eeq
where
\beq
 \ba{rcl}
 a_{\Lambda}^{(m)}&=&
a_{(\lambda^s;\lambda^a)}\, ,\\
&=&\sum_{i_1<i_2<\ldots
<i_m}\theta^{i_1,\ldots,i_m}m_{\lambda^s}(z_{i_1},\ldots, z_{i_m})
a_{\lambda^a}(z/\{z_{i_1},\ldots, z_{i_m}\})
\ea
\eeq
where $a_\lambda$ is the antisymmetric monomial function defined in
\reff{antisimf}.  The eigenvalues are given by:
\beq
\varepsilon_{\Lambda}^\prime=\varepsilon_{\lambda}^\prime=\varepsilon_{-\lambda}=\sum_j [\lambda_j^2-\beta(N+1-2j)\lambda_j] \, .
\eeq

\section{Eigenfunction  generators}\label{genesjack}

It is possible to generate eigenfunctions of $\bar{\mc{H}}$ by  applying some
differential operators directly on Jack polynomials.  Let us consider
the operator $\mc{B}_{\gamma}^{(m)}$, indexed by a positive integer
 $m$ and a partition $\gamma$, given by:
\beq
\ba{c}
\mc{B}_{\gamma}^{(m)}=\sum_{P\in S_N}\theta^{P(1,\ldots,m)}\bar{D}^{\gamma}_{P(1,\ldots,N)},\\
\bar{D}^{\gamma}_{(j_1,\ldots,j_N)}
=(\bar{D}_{j_1})^{\gamma_1}\ldots (\bar{D}_{j_N})^{\gamma_N}. \ea
\eeq In this formula,  $\bar{D}_i$ is the Dunkl operator defined
in \reff{defDunkl}. The partition  $\gamma$ is used to identify
the independent polynomials in $\bar{D}_i$.  The
$\mc{B}_{\gamma}^{(m)}$'s commute with the hamiltonian with an
exchange term and are invariant under the action of the exchange
operators $\mc{K}_{ij}$: \beq [\mc{B}_{\gamma}^{(m)} ,
\mc{H}_{K}]=0=[\mc{B}_{\gamma}^{(m)} , \mc{K}_{ij}] \, .\eeq
These properties ensure that the application of $\mc{B}_{\gamma}^{(m)}$ on the
Jack polynomial
$J_{\lambda}(z;1/\beta)$ gives a solution to the supersymmetric
model,
the solution being indexed by two partitions $\gamma$ and $\lambda$:
\beq
\ba{c}
\mc{A}^{(m)}_{\gamma,\lambda}(z,\theta;1/\beta) \equiv
\mc{B}_{\gamma}^{(m)}J_{\lambda}(z;1/\beta)\, ,\\
\mc{H}_{\kappa}\mc{A}^{(m)}_{\gamma,\lambda}(z,\theta;1/\beta)
=\mc{H}_{K}\mc{B}^{(m)}_{\gamma}J_\lambda=
\mc{B}_{\gamma}^{(m)}\mc{H}_{K}J_{\lambda}=\varepsilon_{\lambda}
\mc{A}^{(m)}_{\gamma,\lambda}(z,\theta;1/\beta)\, .
\ea
\eeq
In the previous equation, we have used the equivalence
between the supersymmetric
hamiltonian
$\mc{H}\equiv \mc{H}_\kappa$ and the hamiltonian with an exchange term
 $\mc{H}_K$ when they
act on a function which is invariant
under the operators $\mc{K}_{ij}$.  The solutions
$\mc{A}^{(m)}_{\gamma,\lambda}$ are of degree $(m,|\lambda|)$ in
$z$.  We can easily check that these symmetric superfunctions
are non-zero only if the partitions $\gamma$ are strictly
decreasing with respect to their $m$ first parts.

The operators $\mc{B}_{\gamma}^{(m)}$ can thus generate any $\mc{H}_\kappa$ eigenfunctions.
They play the double role of fermionic creation operators and of symmetrizer in
$\theta$ and in $z$.  Moreover, they generalize the supersymmetric charges introduced in
section \ref{SUSYDunkl}.  For example:
\beq
\Pi_{\mc{K}}(\mc{B}_{(n)}^{(1)})=\bar{\mc{Q}}_{(n)}\, .
\eeq
Given the infinite number of partitions $\gamma$, there exists an infinite
number of such operators, even in a specific fermionic sector.  Obviously,
since the number of Jack superpolynomials is finite for a given degree
 $m$, most of the solutions
 $\mc{A}^{(m)}_{\gamma,\lambda}$ are linearly dependent.   There is however
no precise relation between a given eigenfunction
$\mc{A}^{(m)}_{\gamma,\lambda}$ and the Jack superpolynomials defined
in \reff{sjackensm}
and indexed by a superpartition.

\section{Combinatorics of superpartitions} \label{compteetats}

We now determine the number $p(m;n)$ of independent states of degree
$(m;n)$. This amounts to counting the number of superpartitions
of degree $(m;n)$.
As we will show, this number is simply given by  the following sum:
\beq \label{comptespart}
p(m;n)=\sum_{n_a +n_s=n}p^{(m)}_{\mbox{{\tiny D}}}(n_a) p(n_s)\, ,
\eeq
where $n_a=|\lambda^a|$ and $n_s=|\lambda^s|$.
As usual $p(n)$ is the number of partitions of
 $n$ so that $ p(n_s)$ counts the number of partitions of $\lambda^s$.  The
quantity
$p^{(m)}_{\mbox{{\tiny D}}}(n_a)$
is a restricted  partition whose `restriction symbols' agree with the usual
definitions of combinatorial analysis:
\beq
\ba{l}
p^{(m)}(n)=\mbox{number of partitions of } n \mbox{ of length less or equal to }
m\, ,\\
p_{\mbox{{\tiny D}}}(n) =\mbox{number of partitions of } n \mbox{ whose parts
are all distinct }\, .
\ea
\eeq
For instance, $p^{(2)}(4)=3$ since there are 3 partitions of 4 with at most 2
parts : $(4)$, $(3,1)$ and $(2,2)$.  Hence, $p^{(m)}_{\mbox{{\tiny D}}}(n)$
gives the number of partitions (without zero) of $n$ whose parts are strictly
decreasing and whose length is smaller or equal to
$m$.  This quantity
counts the number of
antisymmetric partitions $\lambda^a$.
The summation takes into account the various ways of splitting $n$ into the
two integers $n_a$ and $n_s$.
Note that, since the superpartitions of the form
$(;\lambda^s)=(\lambda^s)$ are acceptable, we have to adopt the following
conventions:
\beq
p^{(0)}(0)=1 \quad \mbox{ and }\quad p^{(0)}(n\geq 1)=0\, .
\eeq
For example, we find 5 superpartitions of weight 5 in the 2-fermion sector:
\beq
(3,0;0), \quad (2,1;0), \quad (2,0;1),  \quad (1,0;2), \quad  (1,0;1,1)
\eeq
This agrees with eq. \reff{comptespart}:
\beqa
p(2;3)&=&\sum_{n_a +n_s=3}p^{(2)}_{\mbox{{\tiny D}}}(n_a) p(n_s)\, ,\\
&=&p^{(2)}_{\mbox{{\tiny D}}}(3) p(0)+p^{(2)}_{\mbox{{\tiny D}}}(2) p(1)+p^{(2)}_{\mbox{{\tiny D}}}(1) p(2)+p^{(2)}_{\mbox{{\tiny D}}}(0) p(3)\, ,\nonumber\\
&=&2\cdot 1+1\cdot 1+1\cdot 2+0\cdot 3=5\, .\nonumber
\eeqa
If we compare the number of superpartitions of degree
 $(m;n)$
for $n$ fixed but for different values of $m$, we see that this number
is maximum in the one-fermion sector
and minimum in
the
 $N$-fermion sector (since $N$ corresponds to the maximal length of the
superpartition).
Since the antisymmetric partitions are
strictly decreasing, we have: \beq p(N;n)=0 \mtext{if} n\leq
\frac{(N-1)(N-2)}{2}\, . \eeq For the superpolynomials of degree 3 in
 $z$ and  in $\theta$, for instance, only one state exists because there is
only one possible superpartition, $(2,1,0)$.  This is less than
the $p(3)=3$ possible partitions of 3. On the other hand, if the
fermionic sector is $m=1$, we have $p(1;3)$ independent states,
each associated to one of the 7 possible superpartitions: \beq
(3;0), \quad (0;3), \quad (2;1), \quad (1;2), \quad (0;2,1), \quad
(1;1,1), \quad (0;1,1,1)\, . \eeq

\end{appendix}

\vskip0.3cm
\noindent {\bf ACKNOWLEDGMENTS}

We would like to thank Luc Vinet for his collaboration at the initial stage of
this work and for his continuous support and David S\'en\'echal for some
clarifications. P.D. would like to thank the Fondation J.A Vincent for a student
fellowship.  Finally, we acknowledge the financial support of  FCAR
and CRSNG.

\end{document}